\documentclass{emulateapj}
\slugcomment{{\em}}

\shorttitle{Global star formation revisited}
\shortauthors{Silk \& Norman}
\usepackage{hyperref}
\def\simlt{\lower.5ex\hbox{$\; \buildrel < \over \sim \;$}}
\def\simgt{\lower.5ex\hbox{$\; \buildrel > \over \sim \;$}}
\def\simpropto{\lower.2ex\hbox{$\; \buildrel \propto \over \sim \;$}}

\begin{document}

\title{ Global star formation revisited}

\author{Joseph Silk$^1$  \& Colin Norman$^2$ }
\affil{ $^1$Physics~Department,~University~of~Oxford, 1 Keble Road,~Oxford,~OX1~3RH,~UK\\
$^2$Physics~Department, The Johns Hopkins University, 2400 North Charles Street, Baltimore MD 21218}

\email{silk@astro.ox.ac.uk, norman@stsci.edu}

\def\simlt{\lower.5ex\hbox{$\; \buildrel < \over \sim \;$}}
\def\simgt{\lower.5ex\hbox{$\; \buildrel > \over \sim \;$}}
\def\simpropto{\lower.2ex\hbox{$\; \buildrel \propto \over \sim \;$}}

\begin{abstract}

 A general treatment of  disk star formation is developed from a dissipative multi-phase model, with the dominant dissipation due  to cloud collisions. The Schmidt-Kennicutt law emerges naturally for star-forming disks and starbursts. We predict that there should be an inverse correlation between Tully-Fisher law and Schmidt-Kennicutt law residuals. The model is extended to include a multi-phase treatment of supernova feedback that  leads to a turbulent pressure-regulated generalization of the star formation law and is applicable to gas-rich starbursts.  Enhanced pressure, as expected in merger-induced star formation, enhances star formation efficiency. An  upper limit is derived for the disk star formation rate in starbursts that depends on the ratio of global ISM to cloud pressures.  We extend these considerations to the case where the interstellar gas pressure in the inner galaxy is dominated by outflows from a central AGN. During massive  spheroid formation, AGN-driven winds trigger star formation, resulting in enhanced supernova feedback and outflows.   The outflows are comparable to the AGN-boosted star formation rate and saturate in the super-Eddington limit. Downsizing of both SMBH and spheroids is a consequence of AGN-driven positive feedback. Bondi accretion feeds the central black hole with a specific accretion rate that is proportional to the black hole mass. AGN-enhanced star formation is mediated by turbulent pressure and relates spheroid star formation rate to black hole accretion rate. The relation between black hole mass and spheroid velocity dispersion  has a coefficient  (Salpeter time to gas consumption time ratio) that provides an arrow of time. Highly efficient, AGN-boosted  star formation can occur at high redshift.
 \end{abstract}

\hfill\today

\section{Introduction}
Supernova feedback is considered to be a crucial element for negative
feedback in star formation in disk galaxies. The star formation
history in massive spheroids requires, according to the prevalent
view, negative feedback from AGN. Whether this is sufficient to
explain the observed downsizing is far from clear. Here we 
reassess the Schmidt-Kennicutt (SK) star formation law  and
develop a simple multi-phase model in terms of the porosity formalism applied to disk galaxies \citep{sil01}.We extend the model to  incorporate AGN-triggered
star formation and provide an application to spheroid formation and
ultraluminous starbursts.

A cloud collision model of the SK law has been previously presented by
\citet{tan99}, who uses galactic shear to compute the cloud collision
rate. One advantage of this approach is that it provides a natural
explanation for the low star formation rates observed in the outer
parts of disk galaxies and complements an alternative explanation  which appeals to UV background radiation-controlled $H_2$ suppression in the dust-deprived outer disk \citet{sch04}.
  We provide a simplified reformulation below,
that we will apply in the context of a multi-phase medium to
incorporate star formation and supernova feedback (Section 2). In Section 3, we explore 
regulation of star formation by turbulent pressure and set an upper limit on  the disk surface brightness due to star formation. Section 4 builds on the AGN feedback model \citet{sil05} and applies AGN triggering to star formation in protospheroids. Scaling laws are derived for the black hole growth rate and the star formation rate. Downsizing of both super-massive black holes and stellar mass is  found to be a natural consequence of Bondi accretion-fed black holes and AGN-induced star formation

\section{Disk star formation rate: cloud collision model}

Consider cloud collisions in the disk as a trigger of star formation.
Cloud formation and collisions are driven by the non-axisymmetric
gravitational instability of a cold self-gravitating gas-rich
disk. Let a typical cloud have pressure $p_{cl}$ and surface density
$\Sigma_{cl}.$ We expect star-forming clouds to be marginally
self-gravitating and also to be confined by ambient gas pressure.
Clouds  form this way, and may be maintained if the cloud covering factor 
is of order unity,  this condition guaranteeing  that collisions occur on a local dynamical time-scale.
If the clouds are strongly bound, it is difficult to avoid a short lifetime, collapse and star formation. Our description is a statistical one where we are assuming a steady state ensemble of clouds although the clouds are being formed and reformed all the time in competition with cloud destruction and dispersal processes such as star formation and collisions. For typical parameters in the Galaxy, the disk crossing time normal to the disk and the cloud lifetimes are similar both of order $10 \rm Myr,$ although we assume in general a statistically steady state cloud population in this analysis.

The following relations then apply:  
\begin{eqnarray}
p_g  =\rho_g\sigma_g^2=\pi G \Sigma_g \Sigma_{tot}, \end{eqnarray} 
assuming equal scale-heights for clouds and stellar mass. If the clouds are self-gravitating,  then 
$p_{cl} ^{sg}= \pi \chi G {\Sigma_{cl}}^2 , $
where $\chi\sim 10$ is an estimate of the pressure enhancement due to self-gravity of interstellar clouds.
We redefine $p_{cl} ^{sg}=\chi p_{cl} ,$
and can now write
$\Sigma_{cl} = (p_{cl}/p_{g})^{1/2}(\Sigma_{tot} \Sigma_g)^{1/2}.$
The covering factor $S_{cl}$ of clouds in the disk is directly inferred to be 
$ S_{cl}= {\left(\Sigma_g / \Sigma_{cl}\right)}f_{cl} ,$
where $f_{cl} $ is the gas fraction in clouds. We rewrite this as 
$ S_{cl}= f_{cl}{\left(p_{g}/p_{cl}\right)}^{1/2}{\left(\Sigma_g / \Sigma_{tot} \right)}^{1/2}.$ Here 
$\Sigma_{g}$ is the total (cloud plus diffuse) gas surface mass density.
The cloud collision time-scale is
$t_{coll}=({\Sigma_{cl}H})/({\Sigma_{g}f_{cl}\sigma_g}),$ where the scale
height $H^{-1}=({\pi G\Sigma_{tot}})/{\sigma_g^2}$,  
and $\sigma_g$ is the cloud velocity
dispersion.
The collision time can also be expressed as $t_{coll} = {S_{cl}}^{-1} t_{cross}$
with $t_{cross} = {H/{\sigma_g}},$
which becomes
$ t_{coll} =f_{cl}^{-1}(p_{cl}/p_{g})^{1/2}{\left( {\Sigma_{tot} / \Sigma_g}\right)^{1/2} \left(H / \sigma_g\right)} .$
More generally, inclusion of more realistic 3D cloud kinematics (cf. \citet{tas08}) yields correction factors of order unity.

We now assume the disk star formation rate is self-regulated by supernova feedback which drives the cloud velocity dispersion.  While this assumption has  a long history (c.f. \cite{firm92}), it remains controversial. Numerical simulations certainly demonstrate that supernovae provide negative feedback into star-forming clouds by driving turbulence  
\citep{jou06,tas06,koy08a, koy08b, kim07, jou08}.
Turbulent pressure plays an important role in regulating star formation, via controlling the porosity of supernova remnant-driven bubbles \citep{sil01} as well as the molecular hydrogen fraction 
\citep{bli06}.
At the same time,
global shear also plays a role in controlling cloud peculiar velocities, especially for massive clouds
\citep{gam91}.
Since global gravitational instabilities ultimately drive cloud formation, and hence control star formation, the common origin of competitive turbulence drivers means that effects of shear and supernovae in self-regulating cloud turbulence are not easily separated in 2-dimensional models \citep{she08}. However fully three-dimensional high resolution models of self-consistent star-forming disks embedded within dark halos demonstrate that non-axisymmetric gravitational instabilities dominate the observed turbulence  of $\sim 10 \rm km/s$ at low star formation rates, but that supernova feedback will be important via the intermediary of the hot gaseous phase  at a star formation rate in excess of $10^{-3} \rm  M_\odot kpc^{-2}yr^{-1}$ \citep{age09, tam09}.

Let $m_{SN}$ be the mass in stars formed in order to result in a Type II supernova. 
This is just a  function of the adopted IMF.
Momentum balance gives
\begin{eqnarray}\dot\Sigma_\ast({E_{SN}}/({m_{SN}v_c}))=
f_c{\Sigma_g\sigma_g}/{t_{coll}}.\end{eqnarray}
Here $f_c$ is the cloud volume filling factor, which can be expressed in terms of porosity $Q$ as 
$f_c=e^{-Q}.$
Also, $E_{SN}$ is the kinetic energy of a SNe II
and $v_c$ is the velocity at the onset of strong cooling of the SNe II remnant. Canonical numbers used throughout are $m_{SN} = 150 \rm M_{\odot}$ (for a Chabrier IMF) and $v_c= 400\, \rm kms^{-1}$.
 
We can rewrite the star formation rate per unit volume  as 
\begin{eqnarray}
\dot\rho_\ast=\epsilon_{SN}f_c f_{cl} \sqrt{G\rho_g}\rho_g 
\label{eq:AB}
\end{eqnarray}
with $\epsilon_{SN}=({m_{SN}v_c\sigma_g}){E_{SN}}^{-1}({p_g}/{p_{cl}})^{1/2}.$ 
This formulation is commonly used as a star formation rate prescription in semi-analytical modeling of galaxy formation.  It may be more relevant to rewrite this formulation for disks:
\begin{eqnarray}
{\dot{\Sigma_*}} = f_c f_{cl}G \left(\pi \Sigma_{tot}\right)^{1/2}\left({m_{SN} v_c )\over {E_{SN} }}\right)\left[{p_g / p_{cl}}\right]^{1 \over 2} {\Sigma_g}^{3/2} \ \ \cr
= \epsilon_{SN}f_c f_{cl}\sqrt{f_g }
(R/H)^{1 \over 2}\Sigma_{gas} \Omega \ \ \ \ \ \ \ \ \ \ \ \ \ \ \ \
\end{eqnarray}
where
$\Sigma_{tot} =\Sigma_{g} +\Sigma_\ast.$
Here the disk gas fraction is $f_g \sim 0.1$,  and we use the disk scale-height -to-radius relation
$H/R=(\sigma_g/v_r)^2$ for a disk rotating at $v_r$ with $\Omega^2=G\Sigma_{tot}/R.$
Remarkably, although 
the preceding formula ignores  the multi-phase nature of the interstellar medium and the possibility of gas outflows (see below), one nevertheless manages to fit the 
Schmidt-Kennicutt relation.

We write the observed Schmidt-Kennicutt (SK) law as
$\dot\Sigma_\ast=C_{SK}\Sigma_{g}^{3/2},$ and obtain the SK law coefficient 
\begin{eqnarray}C_{SK}=
{\pi}^{1/2} G f_c f_{cl}({m_{SN} v_c} / {E_{SN}} )[{p_g / p_{cl}}]^{1 \over 2}\Sigma_{tot}^{1/2}.
\label{eq:AA}
\end{eqnarray} Inserting typical parameter values, we find that 
\begin{eqnarray}
{\dot \Sigma_\ast} \approx 0.02 \left(\epsilon_{SN} \over 0.02\right)\left(f_c \over 0.3\right)f_{cl}f_g\left(0.1R\over H\right)^{1\over 2}\Sigma_{g}\Omega .
 \end{eqnarray} 
 This demonstrates that we get the correct normalization at, say, 3 kpc, the scale length of the molecular gas
 in the Milky Way,
 where the scale-height is around 100 pc, the gas fraction is around 0.2, and the molecular gas covering fraction  around 30\%. The observed star formation efficiency in inner spiral disks is found to be fairly robust  and for $H_2$ alone amounts to $5.25 \pm 2.5 10^{-10 } \rm yr^{-1}$ \citep{le08}.

This compares well with the Kennicutt law, both locally and at $z\sim 2$  in shape ($\dot\Sigma_\ast \simpropto \Sigma_{gas}^{3/2}$) and in normalization (for $ \epsilon_{SN} \approx
0.02$) \citep{bou07}.  For the luminous starbursts at $z\sim 2, $ the turbulence is enhanced 
($\sigma_g\sim 40\rm km/s $), but the scale height is thickened.  One reason is that $\epsilon_{SN}\propto\sigma_g$ and $(R/H)^{1/2}\propto 1/\sigma_g$ for disks  with varying amounts of turbulence, 
as might be induced by minor mergers. 
If the covering factor increases,
it is not obvious if the star formation rate in a cloud collision model would increase. To lowest order, these effects all cancel at fixed $\Sigma_{tot}$,
and we can hence understand how starbursts remain on the local Schmidt-Kennicutt law.
Supernova feedback effectively keeps star formation inefficient. 

Of course, we need to better understand how starbursts satisfy the same scaling law as quiescent disks. One hint is that while the gas velocity dispersion may vary in starbursts depending on the merging history, $\Sigma_{tot}$ satisfies Freeman's law and is approximately constant for star-forming disk galaxies.
The observed dispersion in the Schmidt-Kennicutt law may arise from the dispersion in total surface density and molecular as well as total
gas fraction $f_g$.

\subsection{Tully-Fisher relation}
The Tully-Fisher relation is also controlled by disk surface density.
We use the empirical I-band Tully-Fisher (TF) relation:
$L_\ast=C_{TF}v_r^\alpha,$ with $\alpha\approx 4$ in the K-band \citep{mas08}
and $v_r$ the maximum rotation velocity,
and where the virial theorem requires that $C_{TF}=(3/{4 \pi})G^{-2}\Sigma_{tot}^{-1}(L_\ast/M_{tot})$.
We find using equation (\ref{eq:AA}) that
\begin{equation} 
C_{SK}
= {3 \over 4}C_{TF}^{-1/2} f_c f_{cl}\left({m_{SN}v_c}\over {E_{SN}}\right)\left[{p_g / p_{cl}}\right]^{1 \over 2}
\left({L_{\ast} \over M_{tot}}\right)^{1/2}.
\end{equation}
We   infer that the Schmidt-Kennicutt law residuals should anti-correlate with the Tully-Fisher law residuals.  
The Tully-Fisher normalization is correct, by assumption: what is new is the predicted inverse correlation between SK and TF law residuals.

\subsection{Gas-dominated disks}
The global star formation law can be applied to regions that are gas-dominated.  When gas dominates the self-gravity, the cloud collision model suggests that 
$\dot\Sigma_\ast \propto { \Sigma_g}^{3/2}{\Sigma_{tot}   }^{1/2}\simpropto {\Sigma_{gas}   }^2, $ and the KS law steepens. There are indications of such a steepening in several environments.
\begin{enumerate}
\item
The extended HI spiral structure in NGC 6946 \citep{boo07} shows that global gravitational instability is not a sufficient condition for forming stars. 
In the case of M83, the HI disk  extends to more than twice the optical scale. Deep UV imaging reveals very low level star formation in the outer HI disk, well below the SK threshold. The cloud collision model provides a possible explanation of these phenomena, although the observed radial dependence of star formation rate is too steep to be explained by the simplest models \citep{le08}.

\item
 Individual young star complexes in M51 fall on the SK law, although with increased dispersion
\citep{ken07} and a slightly steeper slope.  
\item
Damped Lyman alpha systems at $z\sim 2$ \citep{wol06}
underproduce stars by up to a factor of 10 in star formation rate as predicted from  the SK law. 
\item
Steepening also occurs in the inner regions of disks at extreme star formation rates.
This is found in intensely star-forming galaxies at high redshift \citep{gao07}. \citet{kru07} account for the linear relation found for the local HCN data in terms of the  critical density for excitation of the HCN transition in dense gas, that effectively samples only the densest molecular clouds and thereby bypasses the sensitivity to dynamical time-scale.

\end {enumerate}
Steepening in a cloud collision model  is not a unique explanation
for any of these phenomena. For example, the outer parts of disks are more thermally stable \citep{sch04}, and the  star formation rate in DLAs could be suppressed  because of the  low H$_2$ content due to a combination of a low dust content plus a  high radiation field.

\section{Pressure-regulated star formation  and starbursts}
Turbulent pressure-regulated star formation is especially likely to be important in starbursts.
In disks, atomic  cooling provides an effective thermostat for the turbulent velocity dispersion.  Feedback operates via
the hot phase venting into the halo. Gas may cool and fall back into the disk, as in the galactic fountain model, or escape in a wind, as happens for dwarf starburst  galaxies. The simple porosity description of supernova feedback in a multi-phase ISM \citep{sil01} provides an expression for the star formation rate in which porosity-driven turbulence is the controlling factor: 
$\dot\rho_\ast=Q m_{SN}({4\pi/ 3}R_a^3t_a)^{-1},$ where the shell reaches a final size $R_a$ before break-up,  determined by the ambient pressure at expansion time $t_a$.
The shell evolution  is generally described by \citep{cio88}
\begin{equation}
t=t_0E_{51}^{3/14}n_g^{-4/7}(v_c/v)^{10/7} 
\end{equation}
and
\begin{equation}
R=R_0E_{51}^{2/7}n_g^{-3/7}(t/t_0)^{3/10} ,
\end{equation}
where
$v_0=413 \rm km/s,  \ \ \  R_0= 14\rm pc ,  \ \ \ t_0=1.3\times 10^4 \rm yr .$
Here cooling becomes significant  at shell velocity $v_c=413 E_{51}^{1/8}n_g^{1/4}\lambda^{3/8} \rm km\,s^{-1}$ where the  cooling time-scale within a SN-driven shell moving at velocity $v_c$ is 
$t_c=v_c/\lambda\rho$,
\begin{equation}
\lambda^{-1}=3m_p^{3/2}k^{1/2}T^{1/2}/\Lambda_{eff}
\end{equation}
and $\Lambda_{eff}(T)$ is the effective cooling rate ($\simpropto t^{-1/2}$ over the relevant temperature range $10^4<T<10^6 K$ associated with cooling shock velocities $< 100kms^{-1}$ ).

The SNR expansion is limited by the ambient  turbulent pressure to be $\rho_g\sigma_g^2$, and we
identify  $v_a$ (shell velocity at time $t_a$) with $\sigma_g.$
We obtain
\begin{equation}
\dot\rho_\ast=Q\sqrt{G\rho_g}\rho_g({\sigma_g}/{\sigma_{fid}})^{19/7},
\end{equation} where
\begin{eqnarray}
\sigma_{fid}=({c_0G^{1/2}m_p^{3/2}v_0^{19/7}E_{51}^{62/49}}
{m_{SN}^{-1}
)^{7/19} n_g^{-1/14}}
\\
\approx 20 n_g^{-1/14} m_{SN,100}^{-0.37}E_{51}^{0.
47}\rm km/s
\end{eqnarray}
and
$c_0={4\pi\over 3}R_0^3t_0.$ 

The dependence of star formation rate on turbulent velocity is reminiscent of
Barnes's model for star formation in  the Mice galaxies, an ongoing merger \citep{bar04}.  
A turbulence prescription is required to reproduce the observed spatially extended stellar distribution, which is inconsistent with the simple density-dependent Schmidt-Kennicutt law. 

We apply the cloud collision model of Section 2 to compute $Q$.
The star formation rates derived via porosity and via cloud collisions  can  be set equal.
Comparison with the star formation rate
derived from cloud collisions yields
\begin{equation}
Q=\left(\frac{\sigma_{fid}}{\sigma_{g}}\right)^{12/7}f_c f_{cl}\frac{m_{SN}v_c\sigma_{fid}}{E_{SN}\pi^{3/2}}\left(\frac{p_g}{p_{cl}}\right)^{1/2}  ,
\end{equation}
then using $f_c=e^{-Q}$, we find
\begin{equation}
Q e^{Q} = f_{cl}\epsilon_{SN, fid}\left(\frac{\sigma_{fid}}{\sigma_{g}}\right)^{12/7} .
\end{equation}
 This is an explicit expression for the porosity as a function of the turbulent velocity. The star formation efficiency is evaluated here at the fiducial velocity dispersion so that:
\begin{equation}
\epsilon_{SN, fid} =({m_{SN}v_c\sigma_{fid}}){E_{SN}}^{-1}({p_g}/{p_{cl}})^{1/2}.
\end{equation}

There are two regimes: $Q<<1$ and $Q>>1,$ where approximate solutions can be found and the real solution joins them smoothly. For canonical values, with $p_{cl} \sim p_g$ and $\sigma_g \sim \sigma_{fid}$ the right hand side of the above equation for Q  is small (due to the $1\%$ efficiency of star formation calculated previously) and is of order $\epsilon_{SN, fid} \sim 10^{-2}$ and thus 
\begin{equation}
Q \sim f_{cl}\epsilon_{SN,fid}\left(\frac{\sigma_{fid}}{\sigma_{g}}\right)^{12/7}.
\end{equation}
For completeness we give here the case where $Q>>1,$ namely
\begin{equation}
Q \sim  \ln\left[{f_{cl}\epsilon_{SN, fid}\left(\frac{\sigma_{fid}}{\sigma_{g}}\right)^{12/7}}\right].
\end{equation}
An approximate formula encompassing both regimes $Q>>1$ and $Q<<1$ is 
\begin{equation}
Q \sim \ln\left[1 + {f_{cl}\epsilon_{SN, fid}\left(\frac{\sigma_{fid}}{\sigma_{g}}\right)^{12/7}}\right] .
\end{equation}

Generally, we find that the star formation efficiency is 
\begin{eqnarray}
\epsilon_{SN} = Q^{-7/12} e^{-7 Q/12} {\left(\epsilon_{SN, fid}\right)}^{19/12} f_{cl}^{7/12}
\end{eqnarray}
A desirable feature is that the star formation rate vanishes at very large $Q.$ and becomes larger for small $Q$. There is no minimum in the star formation efficiency-as a function of Q but  the above features 
suggest that $Q$ asymptotically becomes constant, of order unity, and self-regulation occurs since as Q exceeds unity it depends only logarithmically on the velocity dispersion. One can better understand why starbursts  lie on the same Schmidt-Kennicutt law if we assume that local physics specifies the gas fraction converted into stars, in effect $\epsilon_{SN},$ as in the model of \citet{kru06}. This is plausible for  individual molecular cloud complexes. A constant star formation fraction (equivalently, efficiency) is also expected globally in quiescent disks. Since
$v_{turb}$ self-regulates at $\sigma_g\sim 10\rm km/s, $ then  if porosity also self-regulates, the efficiency or fraction of gas converted into stars per dynamical time is constant and small. Then, the higher turbulence in a starburst means that the porosity is low. In turn, low porosity guarantees  inefficient feedback and runaway star formation.

In merger-driven starbursts, the porosity is small since $\epsilon_{SN}\simpropto\sigma_g,$
whence $Q\propto \sigma_g^{-12/7}.$  Small $Q$ suggests  a nuclear starburst, whereas large $Q$ regulates  global feedback and the star formation rate in a disk. 
This is complicated by the dependence of  $\epsilon_{SN}$ on $v_{turb}$ which is compensated by the increase in scale-height with enhanced turbulence.

In a  
quiescent star-forming disk galaxy, we might expect the porosity to self-regulate and be of order unity. This is the case, for example for the Milky Way, where the supernova feedback regulates the global star formation rate. Even in this case, the star formation is not monotonically decreasing with time, as the Schmidt-Kennicutt law would suggest. Numerical simulations \citep{sly05} suggest that mini-starbursts occur stochastically, on a scale of order 1 kpc, with the 
mean global value  decreasing as the gas supply is reduced.  
Observational evidence for a non-monotonic star formation history in the solar neighborhood comes from surveys of chromospheric age indicators \citep{roc00}. Evidence for a series of starbursts 
is found in the inner disk  star-forming regions of spiral galaxies \citep{all06}.

 Another aspect is the extreme pressure induced by gas dissipation.  This must play a role in regulating star formation.
An  explicit case for pressure regulation is made by 
\citet{bli06} who propose
a modified Kennicutt law:
$\dot\Sigma_\ast\propto\Sigma_g p_g^{0.9}.$
Our SN-regulated law using equation (\ref{eq:AB}) is
\begin{eqnarray}
{\dot\Sigma_{\ast}}= G^{1/2}({m_{SN} v_c / E_{SN}})[{p_g / p_{cl}}]^{1 / 2}{p_g}^{1 / 2}\Sigma_g ,
 \end{eqnarray}
becoming 
\begin{eqnarray}
\dot\Sigma_\ast = \pi^{-1/2}{m_{SN} v_c (E_{SN} \Sigma_{cl})^{-1}}p_g \Sigma_g .
\end{eqnarray}
This is close to a pressure-regulated star formation law, if all star-forming clouds have a threshold column density.

\subsection{Outflows}
If the porosity is high, as 
 may happen transiently in starbursts, we assume that disk outflows occur. These may be winds from dwarf galaxies or fountains in the case of more massive disks. Numerical simulations of star formation in the multi-phase interstellar medium of a disk galaxy are able to model the disk outflows and global star formation history \citep{tas06}. It is useful however to provide an analytic formulation.
Let   $f_L$ be the hot gas loading factor.  It is measured to be around 10 for the outflow from NGC 1569 \citep{mar05}.

Now  the outflow from the disk is
\begin{eqnarray}
\dot M_{out}=(1-e^{-Q})f_L \dot M_\ast   \approx  Q f_L \dot M_\ast .
\end{eqnarray}
This tells us that 
$
\dot M_{out} / \dot M_\ast=f_L Q\simpropto 
\sigma_g^{-1.7}.
$ Low porosity suppresses outflows so that outflows are  suppressed in massive potential wells.

We can also express the outflow rate as 
\begin{eqnarray}
\dot M_{out}\approx Q^2f_L({\sigma_g}/{\sigma_{fid}})^{2.7}f_g^{1/2}M_g/t_d. 
\end{eqnarray}
This  shows that outflows could indeed occur from massive potential wells if porosity can somehow be maintained. We argue below that AGN-triggered star formation fulfills this role.

Outflows are important for dwarf galaxies, but are seen to be quenched in deep potential wells as well as at extreme porosity. 
If $\sigma_g$ is high, the porosity is low, feedback is suppressed and supernova-driven winds are quenched. Even if  $\sigma_g$ is low,  outflows may be suppressed if the gas is dense, leading to low porosity. 
Dwarf galaxy outflows play an important role in IGM enrichment at high redshift.
The mass ejected  during  dwarf formation is comparable to the mass retained in stars formed. Nearby starbursts display this trend, suggesting that the effect may be generic to  powerful starbursts 
in dwarf galaxies.  

\subsection{Upper limit on the disk star formation rate }

\citet{meu07} and \citet{hat08} report a bolometric  upper limit on the disk surface brightness in starbursts of $2\times 10^{11}\rm L_\odot kpc^{-2}$ over $0.1\simlt R_e \simlt 10\rm kpc.$ We interpret this upper limit on disk luminosity  for our model galaxy in terms of radiation pressure and mechanical pressure from limiting the gas surface density ({cf.} \citet{tho05}. 

To avoid  lift-off via application of the Eddington condition requires:\\
(1) for radiation pressure:
\begin{equation}
\Sigma_L <(\pi/2) G c \Sigma_{tot}\Sigma_g
\label{eq:AC}
\end{equation}
or $\dot\Sigma_\ast <c\pi Gf_l^{-1}\Sigma_{tot}\Sigma_{g}/2,$
where $f_l=\epsilon_l c^2\approx 10^{-3}-10^{-4}c^2$ for  
massive stars. Let us evaluate this expression after using the earlier expression for ${\dot \Sigma_{\ast}} $ to eliminate $\Sigma_{g}.$ 
We find using equation (\ref{eq:AC}) that star formation is reduced at low gas surface density and lift-off can be avoided  if
\begin{eqnarray}
 \left[ \left({\epsilon_l m_{SN}  c^2 \over E_{SN}}\right)\left({v_c \over c}\right)\right]^2  \left[{p_g \over p_{cl}}\right]\leq {\Sigma_{tot} / \Sigma_g} .
\label{eq:AD}
\end{eqnarray}
(2) For mechanical energy input: we have instead no lift-off provided that
\begin{eqnarray}
\left[\left({\epsilon_{m} m_{SN} c^2 \over E_{SN}}\right) \left( {v_c \over v_w}\right)\right]^2 \left[{p_g \over p_{cl}}\right] \leq {\Sigma_{tot} \over \Sigma_g} .
\label{eq:AE} 
\end{eqnarray}
Here the symbol $\epsilon_m$ denotes the efficiency of conversion of rest mass energy to mechanical energy in massive stars.
For the lift-off limits to be reached the left-hand side of the above two equations must be greater than unity since $\Sigma_g \leq \Sigma_{tot}$.
For the radiation case, unity is not normally reached on the left hand side unless some of the parameters such as $\epsilon_l$ are significantly different from their normal values. This is what led \citet{tho05} to argue for an AGN luminosity driving this Eddington limit for disks. For mechanical winds, the left side is boosted over the equivalent radiation driving term by a factor of $(c/v_w)^2$ making it reasonable that such mechanically driven feedback occurs. For completeness and because of our poor knowledge of star formation at high redshift we will keep both the radiation and mechanical estimates in our calculations.       
This  upper limit on    $ \Sigma_g$ 
corresponds to an upper limit on global star formation of $45 \rm M_\odot kpc^{-2}yr^{-1}.$ 
By writing 
the disk surface brightness as $\Sigma_L=\epsilon_l c^2\dot\Sigma_\ast,$ 
we find using equation (\ref{eq:AD}) that the disk surface brightness satisfies
\begin{eqnarray}
\Sigma_L<(\frac{\pi}{2})^2 Gc\Sigma_{tot}^2\left[\frac{E_{SN}}{m_{SN} c^2\epsilon_l}\left({c \over v_c}\right)\right]^2 \left(\frac{p_{cl}}{p_{g}}\right) .
\label{eq:AF}
\end{eqnarray} 
For mechanical input we have a similar equation using equation (\ref{eq:AE})
\begin{equation}
\Sigma_L<(\frac{\pi}{2})^2 Gc\Sigma_{tot}^2\left[\frac{E_{SN}}{m_{SN} c^2\epsilon_m}\left(v_w \over v_c\right)\right]^2 \left(\frac{p_{cl}}{p_{g}}\right) .
\label{eq:AG}
\end{equation} 
We see that the absolute maximum global star formation-driven surface brightness, $ \Sigma_{L, max}$, is 
\begin{equation}
\Sigma_{L, max}=(\pi/2)Gc \Sigma_{tot}^2.
\end{equation}
This expression can be derived directly from the previous equation
by setting $\Sigma_g = \Sigma_{tot}$. Note in this case all the photon energy is used up to support the disk and the photon bubbles can occur with the signature of light and dark patches on the disk.
This suggests saturation of the star formation rate occurs at high surface density. Inserting typical numbers, 
$\Sigma_L<10^{11}\rm L_\odot \, kpc^{-2} E_{51}^2v_{c,400}^{-2}\epsilon_{n,-3}^{-2}\Sigma_{tot,1000}^2$.
Note that \citet{kom05}  report an offset roughly equivalent to  an effective steepening of the SK law for ultra-luminous starbursts, as do \citet{gao07}.
We now find that for an extreme starburst at the Eddington luminosity with Thomson scattering, $\sigma_T$, as the dominant opacity the relation between surface brightness and mass surface density is:
\begin{equation} 
\Sigma_L =4 \pi Gc \Sigma_{tot} \left({m_p \over \sigma_T}\right)
\label{eq:AH}
\end{equation}
where $m_p$ is the proton mass. From (\ref{eq:AF}) and (\ref{eq:AG}) we deduce that
 the total surface density, which is essentially all gas in initial situations, satisfies
\begin{equation}
\Sigma_{tot} >\Sigma_\ast \equiv \left(\frac{16}{\pi}\right)\left[\frac{m_{SN}\epsilon_l c^2}{E_{SN}} \left({v_c \over c}\right)\right]^2\left(\frac{p_{g}}{p_{cl}}\right) \left(\frac{m_p}{\sigma_T}\right) \end{equation}
$$\sim 10^3\rm M_\odot \,pc^{-2}.$$
and similarily for mechanical input:
\begin{equation}\Sigma_{tot} >\Sigma_\ast \equiv
\left(\frac{16}{\pi}\right)
\left[\frac{m_{SN}\epsilon_m c^2}{E_{SN}} \left(v_c \over v_w\right)\right]^2
\left(\frac{p_{g}}{p_{cl}}\right) \left(\frac{m_p}{\sigma_T}\right) . \ \ \end{equation}
A direct and simple way to derive these limits is to take equations (\ref{eq:AH})and (\ref{eq:AD}) and find that
\begin{equation}
\Sigma_{tot} \geq 8 \left( m_p \over \sigma_T\right)\sim 4 \times 10^{5} M_{\odot} \rm pc^{-2} 
\end{equation}
  which is a derivation of Fish's Law.  
The densest galactic molecular clouds, traced by $H_2O$ masers, have a similar surface density \citet{plu97}.
 By $z \sim 2$, there is a substantial population of galaxies with star formation rates of 500-1000$M_\odot$/yr, ULIRGS and SMGs. A recent study of SMGs at 0.5 arcsec resolution \citet{tac08} finds that they are gas-rich (molecular gas fraction $\sim 30\%$) with the gas in compact disks at a density of $\sim 10^4 \rm M_\odot \,pc^{-2},$ and are undergoing major mergers.  
 
\section{Application to Active Galactic Nuclei}

AGN outflows over-pressure the interstellar medium. They can
deplete the gaseous environment by driving a wind. AGN outflows are the principal element in semi-analytical modeling of massive ellipticals that helps quench recent star formation. The energetics are as follows:
the specific energy per baryon from supernovae is
$E_{SN}/m_{SN}\approx 10^{-5}c^2\rm ergs/gm, $ whereas AGN outflows provide $\sim 10^{-4}c^2 \rm ergs/gm $ per unit spheroid mass, for an assumed efficiency of $0.1c^2$ and a SMBH-to-spheroid mass ratio of $10^{-3}.$ We argue below that AGN outflows have global impact by driving overpressurised cocoons into the inhomogeneous ISM.

Useful formulae  are:\\
 (1) the Eddington luminosity
\begin{equation}
 L_{Edd}=4\pi c G M_{BH} m_p/\sigma_T,
\end{equation} 
(2) the Salpeter time-scale
\begin{equation}
 t_S=\eta c^2 M_{BH}/L_{Edd}=\eta \sigma_T c(4\pi G m_p)^{-1},
\end{equation}
 and \\
  (3) the self-regulating feedback mass 
\begin{equation} 
M_{SRF}=f_g{\sigma_T\over  m_p}{\sigma^4\over{\pi G^2}}.
\end{equation}

Blow-out  by radiation pressure occurs (assuming  a homogeneous interstellar medium)
if  $L=L_{Edd}$ at
$M_{BH}=M_{SRF}$ \citep{sr98,fab99}.
This will lead to a wind, deplete the baryon reservoir and quench black hole
growth by gas accretion. With $f_g\sim 0.1,$ as expected initially in the
protogalactic core, this simple relation fits the mean of the observed
relation
over 3 orders of magnitude in black hole mass. Numerical simulations generally confirm these
simplistic estimates.
We further define a critical AGN luminosity $L_{SRF}$ for star formation-boosted AGN outflow by the  Eddington luminosity associated with the critical black hole mass $M_{SRF}$
that corresponds to the balance between Eddington luminosity and proto-spheroid self-gravity,
\begin{equation}
 L_{SRF}=L_{Edd}M_{SRF}/M_{BH}= 4\sigma^4 c / (Gf_g).
\end{equation}

\subsection{AGN triggering of star formation}

 Negative feedback helps account for the black hole mass-$\sigma$ correlation
 \citep{dim08} and for the luminosity function of massive galaxies \citep{bow06, cro06}.
 More physics must be added however to account for downsizing and efficient star formation in massive galaxies.   The key may be AGN outflows that can  trigger star formation by compressing dense clouds. 
 These would precede the outflow phase which in this case is due to the combined effect of AGN outflow and triggered SNe. A prior phase of positive  feedback is a possible new ingredient in feedback modeling and is motivated by evidence (admittedly sparse but compelling:
 see e.g. \citet{fea07}) for AGN triggering of star formation.
 
The following model is necessarily schematic pending fully three-dimensional simulations of jet propagation into a clumpy proto-galactic interstellar medium.  We speculate that the triggering works as follows. Jet propagation into a clumpy medium develops into an expanding, over-pressurized cocoon \citep{sax05, ant08}. This adds a potentially large multiplier to the
efficacy of the BH-driven outflow for the following reason. The jet-driven plasma-filled radio lobe drives a cocoon that expands into the hot virialized  component of the proto-galaxy at a speed  $v_{co}$ that is much larger than the velocity field associated with the gravitational potential well.   Proto-galactic clouds that are above or near the Jeans mass will be induced to collapse. 
The cloud over-pressuring and resulting triggered star formation occurs at a rate much larger, by 1 to 2 orders of magnitude, than is associated with normal gravitationally-driven fueling, as would be  appropriate to a star-forming disk galaxy. We show below that the star formation rate  enhancement 
amount to a factor $\sim v_{co}/\sigma.$ Hence central massive black holes that have not yet grown by accretion to the limiting mass controlled by radiation outflow should still have a considerable impact on the evolution of the protogalaxy core. Incidentally, this early phase of black hole feedback  makes observational confirmation difficult, as discussed below.

Numerical studies of radiative shock-induced cloud collapse reveal the complex interplay of 
hot and cold gas \citep{mel02, fra04}. Here we focus on the implications for star formation via analytic considerations.
For a simple ansatz, suppose that the triggered star formation
occurs over the cocoon propagation time, $t_{co}$ that is much less than the dynamical time, $t_d$, 
$t_{co} \ll t_d \, $. We show below that
jet triggering of the cocoon expansion gives a star formation rate  enhancement factor
$ t_d/t_{co}\sim v_{co}/\sigma \, .$ 

Although we shall generally assume $v_{co}$ is a paramenter of the wind, consider the case in which the injected momentum flux is $L/c$ for a wind that is  mechanically driven but is originated via radiation
pressure. We expect the mechanical jet luminosity to be a  fraction $v_w/c \sim 0.1$ of the luminosity.  More precisely, the ratio
$(L_{mech}/v_w)/(L/c)$ is equal to the optical depth $\tau$,
due to a combination of Thomson scattering, line opacity and/or dust.
To estimate $\tau,$ an effective  Rosseland mean opacity can be defined 
from these opacity contributions.
This is valid if the jet is radiation pressure-driven.
Now 
\begin{equation}
v_{co}= \sigma \tau^{1/2} f_g^{-1/2}\left({L^{AGN} \over L_{SRF}}\right)^{1/2},
\label{eq:HK} 
\end{equation}
 where we have set 
$\rho_g=\sigma^2/2\pi G r^2.$  
The previously derived  star formation rate  can be generalized for round systems with velocity dispersion $\sigma$ to
\begin{eqnarray}
{\dot M_{\ast}} =( \epsilon_{SN}/\sigma)f_c f_{cl} M_g(Gp_g)^{1/2}.
\label{eq:AI}
\end{eqnarray}
Thus if 
the gas pressure is replaced by the 
AGN-driven pressure,  we can see how a central AGN  can boost the star formation rate by writing the pressure as
\begin{eqnarray}
p_{AGN} = {L_{AGN}  {4\pi v_{co} r^2}}^{-1}.
\label{eq:AJ}
\end{eqnarray}
The star formation rate is now given generally using equations (\ref{eq:AI}) and (\ref{eq:AJ}) by 
\begin{equation}
{\dot M_{\ast}}^{AGN} \approx  \left[ f_c f_{cl} \epsilon_{SN}\left[M_g \over t_d \right]\right] {f_g}^{-1/2} \left({ c \over V_{AGN}}\right)^{1/2} \left( L_{AGN} \over  L_{SRF}\right)^{1/2} .
\label{eq:HL}
\end{equation}
Here we have left $V_{AGN}$ as an arbitrary variable to be applied to the appropriate situation such as jet, wind, outflow etc. In the radiation-driven case discussed above $V_{AGN} = v_{co}, $ the cocoon velocity, and using equations \ref{eq:HK} and \ref{eq:HL} we find
\begin{equation}
{\dot M_{\ast}}^{AGN} \approx  \left[ f_c f_{cl} \epsilon_{SN}\left[M_g \over t_d \right]\right] {f_g}^{1/4} {\tau}^{-1/4}\left({ c \over \sigma}\right)^{1/4} \left( L_{AGN} \over  L_{SRF}\right)^{1/4} .
\end{equation} 
The AGN-driven enhancement factor is $(p_{AGN}/p_g)^{1/2}\approx (v_{co}/\sigma)\tau^{1/2}.$ 
Note that $\epsilon_{SN}\propto \sigma,$ so that the star formation efficiency coefficient (fraction of stars formed per dynamical time) is boosted considerably for spheroids relative to disks.
The AGN luminosity explicitly drives star formation. It is the AGN-triggered star formation multiplier rather than the AGN itself that drives the feedback. The boost effect is generally important in the innermost spheroid, and  globally important  for super-Eddington AGN luminosities.

Note that we must be careful that the AGN pressure does not blow away the ISM completely.
For a SNe-regulated ISM, the pressure relates to $Q$ and includes details of the SNe bubble evolution. But when the AGN jet provides the pressure
it will over-pressure the clouds and turn them into stars with some efficiency. 
We need to relate the star formation from the triggered 
clouds to the AGN accretion rate and this would then affect the jet luminosity and pressure.
Essentially, we raise the external pressure (here using the AGN) on the star-forming clouds. This allows us to regulate $Q$ for the SN bubbles.  If the porosity is maintained to be constant by the triggering of massive star formation, we have achieved the desired self-regulation between AGN activity and star formation. If  $Q$ is given by other means (and is of order unity), then  star formation-regulated feedback follows naturally.  
To show this point explicitly in our formulation of AGN driven star formation we return to the 
pressure-regulated model of the structure of the ISM, and instead of calculating the final size of 
SNeII-driven remnant bubbles by setting the remnant's velocity, $v_a$ equal to $\sigma_g$ we set the external pressure $ \rho_g {\sigma_g}^2$ equal
to the mechanical pressure driven by the AGN, $L / 4 \pi r^2 v_{co}$. Assuming, as above, that $\rho =  {{\sigma_g}^2 /{Gr^2}}$ with an associated pressure $p_g = {{\sigma_g}^4 /{Gr^2}}$ and replacing this with the AGN pressure, $p_{AGN}$ given by $p_{AGN}= L_{AGN} /{4 \pi V_{AGN} r^2}$ we find that the required transformation to replace gas pressure with
AGN pressure is
\begin{eqnarray}
\left({ \sigma_g \over \sigma_{fid}} \right) =\left({c \over V_{AGN}}\right)^{1/4} \left( {L_{AGN} \over L_{SR, fid}}\right)^{1/4} .
\end{eqnarray}
 It then follows directly that 
\begin{equation}
Q e^{Q} = f_{cl}\epsilon_{SN, fid} \left({V_{AGN}\over c}\right)^{1/4}\left({L_{SR,fid} \over L_{AGN}}\right)^{3/7},
\end{equation}
 and for $Q<<1$  we find          
\begin{equation}
Q \sim f_{cl}\epsilon_{SN, fid} \left({V_{AGN}\over c}\right)^{1/4}\left({L_{SR,fid} \over L_{AGN}}\right)^{3/7} .
\end{equation}
For $Q>>1$ 
\begin{equation}
Q \sim  \ln\left[f_{cl}\epsilon_{SN, fid} \left({V_{AGN}\over c}\right)^{1/4}\left({L_{SR,fid} \over L_{AGN}}\right)^{3/7}\right] ,
\end{equation}
with a good fit to the range $Q<<1$ up to $Q>>1$ given by
\begin{equation}
Q \sim  \ln{\left[1 + f_{cl}\epsilon_{SN, fid} \left({V_{AGN}\over c}\right)^{1/4}\left({L_{SR,fid} \over L_{AGN}}\right)^{3/7}\right]}.
\end{equation}

\subsection{ Cocoon Overpressure and the Bonnor-Ebert Condition}
Following the simple cocoon model of \citet{beg89},  we examine the effect of the power flow in the canonical two bidirectional jets diverted into a small nuclear cocoon and then {\it via} the cocoon pressure acting back on the 
central nucleus where gravitational instability and enhanced collapse and accretion may occur if the Bonnor-Ebert critical pressure is reached. For an isothermal gas distribution with velocity dispersion $\sigma$ as used throughout this paper, a jet opening angle $\Theta_J$, an approximately ellipsoidal cocoon with axes $a$ and $b$ with $a>b$  for simplicity powered by two thin jets with luminosity, $L_J$, we find that the cocoon pressure is given by:
\begin{equation}
P_{co} =  \left({ L_J \sigma^2 v_J \Theta_J \over G }\right)^{1/2} \left( a\over b \right)^2
\end{equation}        
and the ratio of the cocoon pressure to the gas pressure in the central region (putting $a \sim b$) for simplicity is
\begin{equation}
{P_{CO} \over P_g} = \left( L_J \over L_{SRF}\right)^{1/2} \left({ v_J c \over \sigma^2} \right)^{1/2} \Theta_J^{1/2} 
\end{equation}
Now for the central isothermal core of gas pressured by the cocoon pressure, the ratio of the Bonnor-Ebert mass to the core mass is given by 
\begin{equation}
{M_{BE} \over M} = 1.18 \left({L_J \over L_{SRF}}\right)^{-{1/4}} \left({ v_J c \over \sigma^2} \right)^{-{1/4}} {\Theta_J}^{-{1/4}}
\end{equation} 
Thus, if $v_J \sim 0.1c $, the core can be overpressured by the cocoon if $\Theta_J > 10 {(\sigma / c)\sim 10^{-2}}$ when $L_J \sim L_{SRF}$. This may be another way to look at the feedback procees involving the growth of black holes along the $M_{BH}-\sigma$ line since at the Eddington luminosity $L_J \sim L_{SRF}$ implies $M \sim M_{SRF}$.     
\subsection{AGN winds}
 
 The global mass loss for AGN-driven wind is given by
\begin{equation}
\dot M_{out}^{gal}= {L_{AGN}\over V_{AGN}^2}
\end{equation}
and for the radiation driven AGN-generated wind case
\begin{eqnarray}
\dot M_{out}^{gal}= {L_{AGN}\over c v_{co}}\approx \left({{\sigma_g}^3\over G }\right) {\tau}^{-1/2} \left({L_{AGN}\over L_{SRF}} \right)^{1/2}.
\label{eq:HM}
\end{eqnarray}
The outflow rate is proportional to the spheroid velocity dispersion and 
to the square root of the AGN luminosity.
The scaling of the outflow rate with regard to AGN luminosity
 implies that outflows saturate. It
may be compared with observations of broadened features that demonstrate the presence of   massive winds in ultraluminous star-forming infrared and radio galaxies.

Making use of the AGN-enhanced star formation rate, we can express the outflow as a ratio using equations (\ref{eq:HL}) and ( \ref{eq:HM}) generally as: 
\begin{equation}
{\dot M_{out}^{gal}\over \dot M_\ast^{AGN}}= \left( { {f_g}^{1/2} \over f_c f_{cl} \epsilon_{SN} ( c/\sigma)}\right) \left( L_{AGN} \over L_{SRF}\right)^{1/2} \left({ c \over V_{AGN}}\right)^{3/2} . 
\end{equation}
It follows that the outflow rate from AGN is always of order the AGN-boosted star formation rate for the proto-spheroid.
We may also compare the star formation-boosted global outflow rate with the AGN outflow. 
The AGN mass outflow rate is 
$\eta  c\dot M_{acc}/(v_{out}),$ with $v_{AGN}\sim 0.1c.$ Hence it is of order the SMBH accretion rate.
As expected, 
the mass flux associated with the galaxy outflow dominates that from the AGN. The mass flux ratio is
\begin{equation}
{\dot M_{out}^{gal} \over \dot M_{acc}} 
=\eta f_w^{-1} \left({c\over \sigma}\right)\left(L_{SRF}\over L_{AGN}\right)^{1/2}.
\end{equation}  
Hence 
${\dot M_{out}^{gal} / \dot M_{acc}} \sim 100 $ for AGN at the Eddington luminosity and 
$\kappa \sim 10^3 \sigma_T/m_p.$ In order to allow for dust, if a factor $\tau^{-1}$ is incorporated into the definition of $L_{SRF},$ this ratio is seen to be inversely proportional to the square root of the adopted (dust) opacity.
The momentum flux ratio is
\begin{equation}
{\dot M_{out}^{gal} \sigma \over {\dot M_{acc} v_{AGN}}}
=\eta f_w^{-1} {c\over v_{AGN}}  \left(L_{SRF}\over L_{AGN}\right)^{1/2}.
\end{equation}
 We see that
the momentum ejected from the AGN dominates over that in the global outflow by a factor of a few for AGN near the Eddington luminosity $L_{AGN}\sim L_{SRF}.$

\subsection{Downsizing}

The piston model enables downsizing of AGN and spheroids by coupling their growth. 
For some fiducial AGN energy conversion efficiency $\eta$ $(\sim 0.1),$ we note that
$L^{AGN}=\eta  c^2\dot M_{acc}$ is a measure of
the BH accretion rate. Since $L^{AGN}$ controls the star formation rate and is itself controlled by the black hole accretion rate, we infer that  black hole growth and star formation triggering downsize together, provided
 $Q$ is approximately constant due to AGN triggering of SN.
The AGN driving of star formation  overcomes the pressure suppression of porosity in the absence of the AGN. 
A large porosity also results in a wind.
The required turbulent velocity field controls the accretion rate and might be
specified by other physics, such as a  merger, or even  be due to the AGN itself.  Let us try to make these assertions more quantitative.

The AGN is the ultimate driver of the porosity. We need to connect AGN-induced star formation
and outflows to the black hole growth rate via the AGN luminosity. The global outflow rate is 
\begin{equation}\dot M_{out}^{gal}= Q f_L \epsilon_{SN}M_g/t_d.
\end{equation}
 By momentum conservation, this  must equal  the global AGN-boosted outflow rate $L^{AGN}/(cv_c).$ 

\subsubsection{Downsizing for porosity-regulated star formation} 
Incorporating the effects of
porosity-driven star formation means that the outflows must satisfy
\begin{equation}
\dot M_{out}^{gal}= 
Q^2f_L({\sigma_g}/{\sigma_{fid}})^{2.7}f_g^{-1/2}\sigma_g^3/G.
\label{eq:AO} 
\end{equation}
The AGN luminosity is controlled by the accretion rate onto the central black hole,
$ \dot M_{acc}. $  Our next  step is to evaluate the black hole growth rate, $M_{acc}.$ This is the key to explaining downsizing. 

To reproduce the downsizing phenomenon, observed for AGN \citep{has05} and their massive host  galaxies \citep{krie07} to occur almost coevally, we need to understand why massive SMBH and spheroids form before their less massive counterparts.
The required scaling for $L^{AGN}$ or $M_{accr}$  is reminiscent of the scaling found for proto-stellar jets. The magnetically-regulated disk phenomenon plausibly obeys a universal scaling law, that could equally apply to jets and outflows from disks around SMBH.  
The proto-stellar scaling is \citep{moh05} $\dot M_{acc}\propto M^2.$ 
\citet{alle06} find that for the black holes  that power the AGNs in massive ellipticals, the Bondi accretion rate is approximately proportional to the jet power.
The connection with outflows and jets that are magnetically guided by 
the wound-up field in the accretion disk proposed by \citet{ban06}
 is a  generic scaling in their study of proto-stellar jets, 
$\dot M_{wind}^{AGN} =f_w \dot M_{acc},$ with 
 $f_w\sim 0.1, $ for outflows associated with central objects that range from brown dwarfs to super-massive black holes.

The Bondi accretion formula therefore regulates SMBH growth and, implicitly,  outflow. It yields
$ {\dot M_{BH}} = \pi G^2\left({p_g / \sigma^5}\right) {M_{BH}}^2 .$
A simple interpretation of this scaling is that for Bondi accretion, 
 \begin{eqnarray}
 \dot M_{out} /f_w=
 \dot M_{acc} =4\pi (GM/\sigma^2)^2\rho v\propto (\rho/\sigma^3)M^2, \ \ 
 \end{eqnarray}
in combination with adiabatic compression, so that $\rho\propto \sigma^3.$ 
For the AGN case, we write
$\dot M_{acc}=\alpha  M_{BH}^2$ with $\alpha \propto {G^2\rho/\sigma^3}.$
Rewriting this we see that 
\begin{equation}
{ d M_{BH} \over d M_*} = Q f_L {f_w}^{-1}
\label{eq:AN}
\end{equation}
Therefore if $Q$ reaches a self-regulating constant and $f_L$ and $f_w$ are also constant then the black hole and galaxy growth move together on a fixed trajectory in the Magorrian plane as discussed below. {\bf \it  It is this fixed trajectory that forces downsizing.}

\subsubsection{Downsizing for SN energy injection}  
Substituting further the relevant quantities for the case of SN energy input we find for the case without AGN feedback 
\begin{equation}
{d\ln{ M_{BH}} \over d{ \ln{ M_{*}}}} =  {f_g}^{1/2}\left({{\bar t_S} \over t_d}\right) = \left( G f_g \rho {\bar t_S}^2\right) ,
\end{equation}
where 
\begin{equation}
{\bar t_S} = \left({ t_S \sigma \over \eta \epsilon_{SN} c f_c f_{cl} } \right)= \beta t_S .
\end{equation}
Thus the critical parameter determining the logarithmic slope in the Magorrian plane is $ {\bar t_S}$. 
There is a critical density $\rho_{crit} = (G f_g {\bar t_S})^{-2}$  above which black hole growth dominates and below which star formation dominates. This can be rewritten in terms of a critical velocity dispersion if one takes $\rho_{crit}$ to be the density at the edge of the Bondi accretion sphere (the sphere of influence of the black hole, $R_{BH}$) namely  $\rho \sim (M/r^3)\sim (M_{BH} / {r_{BH}}^3) \sim (G^{-3} \sigma^6 {M_{BH}}^{-2})$ giving then an equivalent $\sigma_{crit}$  
\begin{equation}
\sigma_{crit} = \left( {M_{SRF} \over M_{BH} }\right) {f_g}^{1/2} {f_c}^{-1} {f_{cl}}^{-1} \left( {p_{cl} \over p_g}\right)^{1/2}\left({ E_{SN} \over m_{SN} v_c}\right), \ \ 
\end{equation}
which is a satisfying combination of black hole-galaxy and ISM properties.
Continuing to the case with AGN feedback, we find
\begin{equation}
{d\ln{ M_{BH}} \over d{ \ln{ M_{*}}}} =  {f_g}^{1/2}\left({{\bar t_S} \over t_d}\right)\left( {v_{AGN} \over c} \right)^{1/2} \left({L_{SRF} \over L_{AGN}}\right)^{1/2}
\end{equation}
and in the radiation-driven case 
\begin{equation}
{d\ln{ M_{BH}} \over d{ \ln{ M_{*}}}} =  {f_g}^{1/4}\left({{\bar t_S} \over t_d}\right)\left( {\sigma \over c} \right)^{1/2}\tau^{1/4} \left({L_{SRF} \over L_{AGN}}\right)^{1/4} .
\end{equation}
Corresponding expressions for the critical density and velocity dispersion in the AGN case can be readily obtained.
At higher redshifts, galaxy systems can be denser, although much of the physics of the nuclear regions depends on local physics, and thus the dominant black hole growth phase may be more easily entered at higher redshift.
The Bondi accretion formula can be rewritten as 
\begin{equation}
\left({t_{BH} \over t_S}\right) = {f_g}^{-1} \left({ t_d \over t_S} \right)^2 \left( {\eta c \over \sigma}\right)\left( {M_{SRF} \over M_{BH}}\right) .
\end{equation}   
Using the equation balancing inflow and outflow and the equation for the logarithmic slope in the Magorrian plane, we find 
\begin{equation}
\left({M_* \over M_{BH}}\right) \left({ Q f_L \over \beta f_w}\right)= \left({t_S \over t_d} \right)\left( {v_{AGN} \over c} \right)^{1/2} \left({L_{SRF} \over L_{AGN}}\right)^{1/2}  .
\end{equation}
Eliminating $ (t_S / t_d)$ we find 
\begin{eqnarray}
\left({ Q f_L \over \beta f_w}\right)^2 \left({t_{BH} \over t_S}\right)=\left( {\eta c \over \sigma}\right)\left( {v_{AGN} \over c} \right) \left({L_{EDD} \over L_{AGN}}\right)\left({M_{SRF} \over M_*}\right)^2  \ \ \ \ \ 
\end{eqnarray}
and in this case we have used the momentum balance equation with momentum injection from SN balancing dissipation. For the radiation-driven case we find    
\begin{equation}
\left({ Q f_L \over \beta f_w}\right)^2 \left({t_{BH} \over t_S}\right) = \tau^{1/2}\left({{L_{EDD}}^2 \over L_{AGN}L_{SRF}}\right)^{1/2}\left({M_{SRF} \over M_*}\right)^2  . 
\end{equation}
In both the above cases there is no evidence for downsizing even with constant Q. However, if there is another way to deduce $Q$ for these turbulent AGN-driven multi-phase media and the momentum injection for the medium is taken up by the AGN then, as we shall show in the following, downsizing can occur naturally as a consequence of the turbulent ISM properties.

\subsection{Constraints on Evolutionary Tracks}\label{sec-constraints}
There is an obvious but powerful constraint on the  behavior of the evolutionary tracks in the observed mass $\ln[M_{BH}] - \ln[M]$ plane. We first emphasize this concept of tracks is implicit in our model and that there is a flow of points in the mass plane with evolutionary arrows all pointing in the direction of black hole growth.       
The slope $p = d \ln[M_{BH}]/ ln[dM] $ of the track cannot be negative since: (I) black holes can only grow in mass, and (II) galaxies only grow in mass (ignoring their small fractional mass loss). On dwarf galaxy scales the fractional mass loss can be easily incorporated but even there it is much less than of order unity. Therefore, for example, tracks cannot loop back from above the mean line with any slope less then zero after overshooting the mean line on a trajectory originating from below. {\it Therefore any non-pathological track will spend most of its life on a track with a slope close to the mean}. Thus we can assume $p$ is approximately constant, Observationally $p$ is of order unity. We now use this slope as a parameter in our time scale and evolutionary equations and find:  
 \begin{eqnarray}
\left({p \over \beta }\right)^2 \left({t_{BH} \over t_S}\right)=\left( {\eta c \over \sigma}\right)\left( {v_{AGN} \over c} \right) \left({L_{EDD} \over L_{AGN}}\right)\left({M_{SRF} \over M_{BH}}\right)^2.  \ \ \ \ \ \ 
\end{eqnarray}   
For the radiation-driven case we find    
\begin{equation}
\left({p \over \beta}\right)^2 \left({t_{BH} \over t_S}\right) = \tau^{1/2}\left({{L_{EDD}}^2 \over L_{AGN}L_{SRF}}\right)^{1/2}\left({M_{SRF} \over M_{BH}}\right)^2 . \ \ \ \ \  
\end{equation}

\subsection{Why the $M^2$ dependence of accretion?}
We  see that the parameter controlling accretion $\alpha $, which is proportional to the phase space density, and  also measures   the specific entropy $s$ of the initial gas distribution, is specified by the physics of accretion.  Specifically, $\alpha \propto s^{-3/2},$ where $s=kTn^{-2/3}.$ In terms of a polytropic equation of state, $\alpha$ is constant for $\gamma=5/3.$ 

\subsubsection{Hot Phase Entropy}
 In any dissipative multi-phase medium, the entropy cannot be strictly constant.
However to demonstrate that entropy is indeed slowly varying in the hot phase of the system system, consider thermal balance between gravitational accretion heating, which is also proportional to the resulting power in the AGN outflow and associated heating, and atomic cooling. One obtains $c_1v^5/G=\rho^2 r^3 \lambda(v), $ where $\lambda(v)$ represents the cooling rate per atom per unit density. In the range of interest where hot gas dominates the gas pressure
on galaxy and cluster scales,   $\lambda(v)$ is weakly varying,
e.g. $\lambda(v)\propto v $ for thermal bremsstrahlung cooling  at $T\simgt 10^7$K and is approximately constant over $10^6-10^7$K. In fact to avoid fragmentation, a necessary condition for effective central black hole growth \citep{lod06}, one needs to be at $T\simgt 10^6$K and have $\gamma >4/3.$  Above $10^7$K, appropriate to massive galaxies and clusters, $s\propto T^{1/3}$ is found to be slowly varying and this helps account for the central entropy "floor"  in clusters. This results in the maximum accretion rate being in the core. Black hole growth by Bondi accretion indeed requires constant specific entropy flow,
which explains why $\dot M_{acc} \propto M_{BH}^2.$ These arguments should apply on massive galaxy scales where the gas pressure is controlled by the ISM hot gas phase at the outer boundary of the flow.

\subsubsection{Cold Phase Entropy}
Similar arguments apply to a cold phase in a multi-phase medium. Here the cold clouds themselves,
envisaged as bound self-gravitating entities  that move on ballistic orbits, act like massive particles whose dynamics can be described by Bondi accretion.  For supersonic turbulence there are three points to note:\\
(1) the system is highly dissipative, and  so momentum conservation not 
energy conservation is the rule;\\
(2) the density as described in the PDF is essentially dimensionless and only measured in units of the square of the Mach number $({\cal M})^2$;\\
(3) continuous energy and momentum input comes for the central source so that the PDF structure of the medium remains statistically robust. This means there will be the same number of clouds with the same mass function even though there is continuous creation and destruction. If a cloud is dissipated then another cloud will be created to take its place in the ensemble.

Finally, the $M^2$ dependence discussed above occurs over a very wide range of sources and environments both relativistic and non-relativistic and with power sources ranging from proto-stars to micro-quasars to quasars. A clear invariant seems to be supersonic turbulence generated by a central source.   
For the cold supersonic phase $ \rho \sim {\cal M}^2$ and $\sigma \sim {\cal M}$ and thus we expect the phase space density be ${\rho / \sigma^3} \sim {\cal M}^{-1}$. Note that over a wide range of the systems discussed from proto-star to quasars the value of ${\cal M} \sim 3-10$ is appropriate. In addition the cool gas over a wide range of conditions is at similar temperatures.The cool gas is carried in packets (in clouds) whose number density and mass function are properties of the supersonic turbulence and the Mach number. Thus over this wide dynamic range the variation in the parameter $\alpha$ may not be large.    

\subsubsection{Multiple phases}
For cloud populations with different properties their effective temperature 
is associated with the  velocity dispersion of the cloud ensemble, as opposed to the gas kinetic temperature. In this case, 
$s= \sum M_i\sigma_i^2 n_i^{-2/3}, $  where we sum over hot and cold cloud components,
and the Bondi accretion rate for a two-phase medium is now
$ {\dot M_{BH}} =  \pi G^2\sum\left({\rho_i / \sigma_i^3}\right) {M_{BH}}^2 .$

\subsection{From Bondi accretion to star formation}
We now develop the interplay between the Bondi accretion rate parameter $\alpha$ and the 
porosity-driven star formation rate. 
We show that constant $Q$ implies constant $\alpha$, and vice versa. This is the key to understanding coupled downsizing for spheroids and super-massive black holes. 
Define the black hole growth time by $t_{BH}=M_{BH}/\dot M_{acc}=1/( \alpha M_{BH})$.

\subsubsection{Significant downsizing for porosity-regulated star formation}
Using
the above equations (\ref{eq:AN}) and (\ref{eq:AO}) which relate  $\dot M_{out}$ and $\dot M_{acc},$ we find that
\begin{eqnarray}
\left({Q^2 f_L \over f_w}\right)^2\left( \eta c \over \sigma_{fid}\right)\left({t_{BH}\over t_S}\right) 
 \ \ \ \ \ \ \ \ \ \ \ \ \ \ \ \ \cr
=
\left({M_{BH}\over M_{SRF}}\right)^2 \left({M_{BH} \over M_g}\right)\left({\sigma_{fid}\over \sigma} \right)^{31/7}  \ \ \  \ \ \ \ \ \ \ \
\end{eqnarray}
which becomes 
\begin{eqnarray}
\left({Q^2 f_L \over f_w}\right)^2 \left({\eta c \over \sigma_{fid}}\right)\left({t_{BH}\over t_S}\right)
\ \ \ \ \ \ \ \ \ \ \ \ \ \ \ \ \cr
=\left({M_{BH}\over M_{SRF}}\right)^2 \left({M_{BH} \over M_g}\right)\left({L_{SR,{fid}}\over L_{SRF}} \right)^{31/28}.
\end{eqnarray}
Alternatively using the $p$-parameterization discussed in section \ref{sec-constraints} we find
\begin{eqnarray}
\left({Q p}\right)^2\left( \eta c \over \sigma_{fid}\right)\left({t_{BH}\over t_S}\right)
\ \ \ \ \ \ \ \ \ \ \ \ \ \ \ \ \cr
=\left({M_{g}\over M_{SRF}}\right) \left({M_{BH} \over M_{SRF}}\right)\left({\sigma_{fid}\over \sigma} \right)^{31/7} \ \ \  .
\end{eqnarray}
 
Constant porosity therefore guarantees downsizing, since 
\begin{eqnarray}
\left({t_{BH} \over t_S}\right)\propto L_{SRF}^{-31/28} \propto M_{BH}^{-{31/28}} \approx M_{BH}^{-1}.
\end{eqnarray}

For Eddington-limited accretion, $M_{BH}= M_{SRF}$.The ratio of black hole growth time to Salpeter time decreases  with increasing black hole mass at constant $Q$. 
Notice also that 
$$1/\alpha = t_{BH}M_{BH}\propto Q^{-4}M_{BH}^2 M_{SRF}^{-59/28}\propto M_{BH}^{-3/28}.$$
 Hence 
constant porosity also 
favors Bondi accretion since $\alpha \approx $ constant. Since $\sigma \sim 10 \sigma_{fid}$ for a
massive spheroid, we also infer that for Eddington-limited accretion,
the porosity must be of order 10 percent if  the black hole growth time is of order the Salpeter time. Indeed, we can equate these time-scales and infer that  porosity depends weakly on black hole mass,
$Q\simpropto M_{BH}^{-1/3}. 
$ 
In the absence of AGN feedback, the derived star formation law yields a ratio of star-formation time-scale to dynamical time that is proportional to $\sigma_g$ and hence approximately constant, independently of galactic mass. This suggests that in starbursts when AGN play no role, there should be no downsizing, as would be expected if internal processes such as those associated with formation of massive star clusters were to dominate. 

  Also, the Eddington ratio can be written as
$$f_{Edd}\equiv {L^{AGN}\over L_{Edd}}=\eta c\left({\sigma_T\alpha}\over {m_pG}\right)M_{BH}.$$ Hence constant $\alpha$ is consistent with 
the observed trend measured in the Eddington ratio \citep{alo08}. The Eddington ratio is found to be  lower for AGN than for QSOs due to a combination of
reduced host galaxy (and SMBH) mass as well as AGN feeding.

\subsection{Why does porosity self-regulate?}
First, we give a qualitative argument for self-regulation. If the porosity $Q$ is low, the jet is blocked,  and the  turbulent pressure is enhanced. Weak shocks propagate ahead  of the cocoon and squeeze  self-gravitating clouds over the Bonnor-Ebert stability limit. Star formation is triggered and the resulting supernovae  drive up the pressure.  Blow-out most likely occurs of the residual gas. This in turn increases 
$Q \sim 1. $ However dense clouds can now fall in unimpeded by intervening dense cold gas.
The  infall reduces $Q$, drives accretion and resurrects a strong jet.

Now, it is reasonable to assume that the cold phase is defined by a minimum density  $n_{cr}$ set by dissipative cooling  and molecular formation, and that is 
not strongly dependent on metallicity and  ionization fraction either in the molecular \citep{nor97} or atomic \citep{wol03} phases.
The shape of the density PDF is well represented by a log normal distribution \citep{wad07} with two parameters, dispersion of the distribution and its amplitude. When calculating  the filling factor of either hot or cold gas,
one integrates the lognormal PDF  below or above $n_{cr}$. At fixed $n_{cr}$, the  filling factor of either hot or cold gas depends only on one parameter, the  dispersion. In general, for the isothermal case considered here, the dispersion is a linear function of $\ln{\cal{M}}$, for high Mach numbers \citep{kru05}. In the adiabatic case the lognormal PDF dispersion is independent of the Mach number. The lognormal form is retained for supersonic turbulence both with and without star formation \citep{wad07}.
Even with significant feedback and increased Mach number, $Q$ only varies logarithmically with Mach number, and therefore any change of filling factor in systems with developed supersonic turbulence is relatively slow as the turbulence  is increased. In summary,
the volume filling fractions of cold gas $e^{-Q}$  and hot phase  $1-e^{-Q} $ depend 
only logarithmically on the Mach number. Hence the porosity is plausibly constant.
Quantifying this argument we define the hot-phase filling factor in a turbulent supersonic medium to be the volume of the turbulent medium with a lower density than the mean by a factor $ \nu_{h} <1$. Using the 
usual lognormal probability density function (PDF) 
\begin{equation}
f(n) = { 1 \over  \sqrt{2 \pi} \sigma_t n} \exp{[-{{(\ln{n})}^2 \over 2 \sigma_t^2}]}
\end{equation}
where for supersonic turbulence we use
\begin{equation}
{\sigma_t}^2 = \ln\left[{ 1 + \lambda_t {\cal{M}}^2}\right]
\label{eq:AP}
\end{equation}
with ${\cal{M}}$ being the Mach number,  and for the parameter $\lambda_t$ 
we use $\lambda_t = 3/4$ \citep{kru05} for numerical estimates. We define 
\begin{equation}
\nu_h = {n_h \over n_0 \left[{ 1 + \lambda_t {\cal{M}}^2}\right]}
\end{equation}  
where $n_0$ is a reference density.
and we find an expression for $f_{h}$
\begin{equation}
f_{h} = {erfc}\left[{ ln\left[{1 \over \nu_h}\right] \over \sqrt{2} \sigma_t}\right]
\end{equation}
where  often the first term of the asymptotic expansion for $erfc{[x]} = {{\exp{[-x^2]}}/x \sqrt{\pi}}+....$ is a useful guide.   
For the Mach number, ${\cal{M}} = 3-10$ and for the under density parameter of the hot phase relative to the mean we use $\nu_h =0.1$ and find the hot phase filling factor is $\sim 10- 20 \%$. As the Mach number increases,  $f_h$ increases in turn as the width of the PDF increases. Since $f_h = 1- \exp{-Q}$ we find that
\begin{equation}
          Q = \ln{\left[ 1-f_{h}\right]}           
\end{equation}
which for $Q<<1$ becomes
\begin{equation}
          Q = \ln{\left[ 1+ f_{h}\right]}           
\end{equation}
giving 
\begin{equation}
        Q \sim \ln\left[{ 1+ {erfc}\left[ln\left[{{1 \over {\bar \nu}} \over \sqrt{2} \sigma_t}\right]\right]}\right] .
\end{equation}
This shows how $Q$ increases as the Mach number increases.   
Proceeding further, we can generalize our previous expression for $Q,$ including both the competing effects of the star formation and supersonic turbulence, to:
\begin{equation}
Q \sim  \ln{\left[ 1+ {erfc}\left[ln\left[{{1 \over {\bar \nu}} \over \sqrt{2} \sigma_t}\right]\right] + f_{cl}\epsilon_{SN, fid}\left(\frac{\sigma_{fid}}{\sigma_{g}}\right)^{12/7}\right]}. \ \ \ \   
\end{equation}
 Feedback from AGN is now readily incorporated by substituting ${\cal{M}}= (\sigma /\sigma_{fid})^2$ and then making the substitute, as before, for the multi-phase medium under the action of the mechanical and radiative pressure originating in the central source giving
\begin{equation}
{\cal M}_{AGN}^2=  \left({c \over V_{AGN}}\right)^{1/2} \left( {L_{AGN} \over L_{SR, fid}}\right)^{1/2}  
\label{eq:AQ}   
\end{equation}
and thus using equations (\ref{eq:AP}) and (\ref{eq:AQ})  we obtain
\begin{equation}
{\sigma_{t,{AGN}}}^2 = \ln{\left[1 + \lambda_t \left({c \over V_{AGN}}\right)^{1/2} \left( {L_{AGN} \over L_{SR, fid}}\right)^{1/2} \right]}   
\end{equation}
In the AGN case it is now clearly quantified how the AGN pressure is reducing $f_h$ by confining the hot SNR bubbles but on the other hand how the AGN increases the Mach number of the turbulence and therefore broadens the PDF distribution, thus increasing $f_h$.  

\subsection {Star formation rate}
Global gas consumption is dominated by star formation, and locally by SMBH growth.  The two are connected via the piston model and suggest a possible self-regulation loop for both spheroid and SMBH growth. We now show that the time sequence 
underlying the Magorrian relation can be interpreted in terms of the ratio of spheroid to SMBH growth rates.

The star formation (or gas consumption) rate is from equation (\ref{eq:HL})
\begin{eqnarray}
{1\over t_{\ast}} ={\epsilon_{SN}\over t_d}\left({ L_{AGN}\tau\over L_{SRF}}\right)^{1/2}\\
= {\bar \epsilon_{SN}}G{\Sigma_{tot}}\left({ L_{AGN}\over L_{SRF}}\right)^{1/2}\\
\propto p_{AGN}^{1/2},
\end{eqnarray}
where 
$\bar \epsilon_{SN} = \epsilon_{SN}/\sigma$ is a constant depending only on supernova properties and the IMF.
This controls spheroid growth and demonstrates spheroid downsizing via the  AGN-enhanced pressure and associated star formation rate. 
 
Writing the AGN luminosity as  $L_{AGN}=\eta\dot M_{BH}c^2,$ 
the Eddington luminosity can be expressed in terms of the Salpeter time as
${\eta M_{BH}c^2 / t_S}.$  We now rewrite the star formation rate expression and obtain 
\begin{eqnarray}
{M_{BH}\over \sigma^4}=\left(t_S \over t_\ast \right)^2\left({ m_p}\over{\bar\epsilon_{SN}\sigma_T \eta c \Sigma_{tot}}\right)^2.
\end{eqnarray}
The predicted normalization of the Magorrian relation agrees with the local value and slope for canonical  
parameter values ($\Sigma_{tot} \sim \Sigma_\ast ,$  $\eta \sim 0.001, $ $\epsilon_{SN} \sim 0.1).$ Of course, there is  considerable uncertainty due to 
possible variations in the  initial mass function, supernova energy and star formation timescale. 

In fact, the relevant SMBH measure in distant objects is $L_{AGN}$ rather than $L_{Edd}.$ Let us make use of $L_{SRF}$ as a  fiducial luminosity, in effect a proxy for $\sigma^4$.We now rewrite the preceding expressions to obtain
\begin{eqnarray}
L_{AGN} / L_{SRF}=\epsilon_{SN}^{-2} (f_g t_d/t_\ast)^2 \sim ( t_d/t_\ast)^2. 
\end{eqnarray}
We also have for the AGN-boosted star formation rate
\begin{eqnarray}
{\dot M_{\ast}} ^{AGN}\approx \epsilon_{SN}M_g \Omega \left( L_{AGN} / L_{SRF}\right)^{1/2} \sim  M_g \Omega.
\end{eqnarray}
This is of course the optimal rate. We also see that
$\dot M_{\ast}^{AGN} \simpropto \sigma^4.$ This is not inconsistent with the observed dependence of outflow velocity on star formation rate \citep{mar05}.

We may consider the case of a recently detected kiloparsec scale starburst at $z=6.24,$ hosted by a luminous quasar which has spatially resolved [CII] emission as well as  a large reservoir of CO-detected molecular gas \citep{fab09}. Other similar  high $z$ objects, detected in CO,  are  believed to be super-Magorrian \citep{mai07}. This quasar host galaxy  also has a star formation rate of $\sim 1000\rm M_\odot year^{-1}kpc^{-2}, $ an order of magnitude larger than is typical of starbursts without  luminous AGN.  For comparison, Arp 220, a low redshift starburst  hosting an AGN of  luminosity comparable to its starburst,  has a similar surface brightness in star formation but only over a 100 pc scale. It is tempting to infer, admittedly with only two well-mapped examples,  that we may be viewing AGN boosting of star formation, with the phenomenon being greatly magnified at high redshift for the most massive objects.

We infer that  if the SMBH mass is super-Magorrian, then  $t_\ast<t_S$ and $t_\ast<t_d$, and star formation is very efficient. This seems to be the case at high redshift.
The preponderance of data indeed suggests that the local relation becomes super-Magorrian prior to $z\sim 2$ \citep{mcl06, woo08} and indeed persists to  $z\simgt 5$ \citep{mai07}. 
Coevolution of AGN accretion and the co-moving star formation rate densities occurs to $z\sim 2$, but the accretion rate falls off relatively towards higher redshift \citep{silv08}. 
Comparison of 
the cosmic star formation history and AGN  accretion rates in comoving number density as a function of luminosity suggests that  the peak in 
massive black hole growth rate occurs several Gyr prior to 
the star formation peak and that downsizing at $z<1$ is due to diminishing accretion rates \citep{bab07}.  Sub-millimeter galaxies (SMGs)  are an exception. SMGs at $z\sim 2$  contain SMBH that are under-massive relative to the Magorrian relation \citep{ale08}. This is suggestive of triggered star formation, which reduces 
$t_\ast$, and may be appropriate in major mergers that generate  dense central gas environments where porosity feedback is suppressed. Note also that the peak in the major merger rate 
also precedes the peak in cosmic star formation rate \citep{rya08}, and is approximately consistent with the peak in comoving AGN accretion rate density.

\section{SUMMARY and CONCLUSIONS}   
In summary,  a general and robust treatment of  disk star formation is developed from a cloud collision model. The Schmidt-Kennicutt law emerges naturally for star-forming disks. We predict that there is an inverse relation between Tully-Fisher law and Schmidt-Kennicutt law residuals. A multi-phase treatment of supernova feedback  leads to a turbulent pressure-regulated generalization of the star formation law that is applicable to gas-rich starbursts. 
Negative feedback from star formation  occurs in disks under  turbulent pressure regulation. In combination with a cloud collision model, the 
Schmidt-Kennicutt law can be understood in diverse environments, spanning quiescent disks
and starbursts. Enhanced pressure, as expected in merger-induced star formation, enhances star formation efficiency.
An  upper limit is derived for the disk star formation rate in starbursts that depends only on the IMF and on the ratio of global to cloud pressures.  For clouds in approximate pressure with interstellar medium and a local IMF, we infer a limiting gas surface density of $\sim 1000\rm M_\odot \, pc^{-2}.$ 
 
We extend these considerations to the case where the interstellar gas pressure in the inner galaxy is dominated by outflows from a central AGN. The star formation rate is pressure-driven and depends on the excess pressure applied by the AGN outflows. 
 During massive  spheroid formation, AGN-driven winds trigger star formation, resulting in enhanced supernova feedback and outflows.
 Downsizing is predicted to be a consequence of AGN-driven positive feedback.
Our most important results refer to downsizing, for which we provide a new interpretation in terms of Bondi accretion feeding of the central black hole.

The specific accretion rate is proportional to the black hole mass. 
We found that Bondi accretion results in $M_{BH}\propto \sigma^{59/7}Q^4 (t_{BH}/t_S).$ This means that
if porosity  self-regulates to be constant, black hole growth proceeds rapidly until it saturates at the Magorrian relation $M_{BH}\propto \sigma^4$ due to blow-out. Black hole downsizing occurs if $\alpha$ is approximately constant. We clarify this as follows.

There are three specific rates that define our model. The Salpeter  rate $t_S^{-1}$ is constant, the black hole growth rate  is $1/t_{BH}= (1/t_S)(L_{AGN}/L_{Edd}), $ and the star formation rate is $ 1/t_\ast=\bar\epsilon_
{SN}(GL_{AGN}\tau)^{1/2}(c\sigma^4f_g)^{-1/2}.$ 
Hence 
\begin{eqnarray}
{t_{BH}\over t_\ast}={{\epsilon_{SN} t_S} \over{ f_gt_d    }}
L_{Edd}\left(\tau\over{L_{AGN}L_{SRF}}\right)^{1/2}\sim { t_S\over{ t_d    }} \tau^{1/2} 
\end{eqnarray} 
and we infer that 
\begin{eqnarray}
L_{AGN} / L_{SRF}\sim( t_d/t_\ast)^2 \sim (t_S/t_{BH})^2.
\end{eqnarray}
This shows that the black hole growth rate and star formation rate are coupled.
At given $\sigma$, there is a critical AGN luminosity, above which AGN-triggered star formation rates dominates over the black hole growth rate.
This critical luminosity is found to be
\begin{eqnarray}
{L_{AGN}^{cr}\over L_{Edd}}= \eta \tau {\epsilon_{SN}^2\over f_g}{t_S\over t_d}{M_{BH}\over M}{c\over \sigma} \left({t_{BH}\over t_\ast}\right)^2 \sim  \left({t_{BH}\over t_\ast}\right)^2 .
\end{eqnarray} 
At super-Eddington luminosities, AGN-triggered star formation dominates.

The model contains two characteristic luminosities which are functions of $\sigma.$ The Eddington luminosity, if combined with the quenching assumption, scales as $\sigma^4.$ The AGN-triggered star formation luminosity   is ${\epsilon_l}c^2{\bar\epsilon_{SN}} \sigma^2(L_{AGN}\tau)^{1/2}(Gcf_g)^{-1/2}.$ This is proportional to $\alpha^{1/2}M_{BH}\sigma^2 .$  Adopting $M_{BH}\propto \sigma^4$
and $\alpha \propto \sigma^{4/3} $ if $Q$ is constant, we find that $L_\ast ^{AGN}\propto \sigma^{20/3}.$  Hence we again infer a critical Eddington luminosity above which triggered star formation dominates the luminosity of the system. This guarantees efficient star formation for luminous  AGN. Moreover if 
$M_{BH}/\sigma^4$ increases with increasing redshift, as inferred from the downsizing of the black hole growth rate, spheroid star formation down-sizes both in mass and in efficiency.

The  ratio of AGN to star formation luminosity is 
\begin{equation}
{L_{AGN}\over L_\ast^{AGN}}={{\eta\dot M_{BH}}\over{\epsilon_n\dot M_\ast^{AGN}}}\propto
{{\rm d}M_{BH}\over {\rm d }M_\ast}.
\end{equation}
 This yields an arrow of time in the form of tracks in the Magorrian diagram. If for each data point in the Magorrian plane, defined by black hole and bulge mass, one can separate star formation and AGN luminosity, then the ratio gives a vector and hence an arrow of time.
This is of course the instantaneous trajectory of points in the Magorrian plane as viewed at any given epoch. However statistically the vectors should provide the flow of galaxy bulges towards the Magorrian relation.  It would be interesting to construct Magorrian flow diagrams  binned  over several  redshift  ranges. This would provide insight into the cosmological evolution of the flow of points in the Magorrian plane.

AGN-enhanced star formation is mediated by turbulent pressure and relates spheroid star formation rate to black hole accretion rate.
 As the AGN pressure is increased, via jet/cocoon pressure-driven interactions with the ISM, the induced star formation rate is correspondingly boosted.  Downsizing for spheroid formation is a consequence of massive black hole downsizing. The observed relation between black hole mass and spheroid velocity dispersion is obtained, with a coefficient   (the ratio of Salpeter time to
gas consumption time)  that provides an arrow of time. Highly efficient, AGN-boosted  star formation is favoured at high redshift. Outflows are of order the AGN-boosted star formation rate and saturate in the super-Eddington limit.

We end  with some relevant questions for observers that are pertinent to our model. Is AGN activity correlated with the star formation rate? Was spheroid formation and black hole growth coeval and symbiotic? Which came first, if either? Is the reach of the AGN too localized to globally affect star formation? Is or was feedback significant in radio-quiet AGN? Are the youngest radio sources, notably the GPC sources, templates for the earliest stages of AGN feedback, and if so, is there associated triggering of star formation? If star formation is triggered by radio cocoons, where is the evidence for a
post-starburst stellar population in old radio lobes? if the efficient mode of star formation is due to coherent cocoon triggering as argued above, what is the evidence for spatially and temporally  coordinated episodes of star formation in well-studied examples such as the Antennae?  Has  the trigger
of positive feedback in ultraluminous starbursts disappeared, due to a short duty cycle,
 or is it well and truly buried in embedded AGN nuclei? Are the associated outflow rates from  AGN/starbursts of any significance for balancing the baryonic budget of the universe, and if so, where does the enriched debris end up?  If the gas remains in the halos of massive galaxies, as essentially all current semi-analytic simulations predict, why hasn't it been seen? And for the modellers (who are  effectively observers of the computer), how can  AGN feedback simulations possibly play any  predictive role if one has to begin with massive seed black holes of dubious heritage and uncertain fate?  Perhaps our analytic discussion will motivate more realistic recipes and sub-grid physics prescriptions for future generations of feedback simulations.

Many of these questions were inspired by discussions with participants at the Oxford-Catania Workshop in Vulcano, May 2008,   on the Interface between  Galaxy Formation  and AGN. We gratefully acknowledge all of them as well as the role of Vincenzo Antonuccio-Delogu, who   organized  a brilliantly topical meeting in the unforgettable setting of the Aeolian Islands as well as provided an opportunity for us to complete this paper.

 The research of CN was funded in part by NASA grant GO 6-1730-X and a JHU/APL collaborative grant.

\section{APPENDIX:LIST OF SYMBOLS}

\begin{description}
\item[$\alpha$]                                        Bondi accretion rate parameter: phase space density
\item[$ \beta$]                                        factor for black hole growth timescale     
\item[$C_{SK}$]                                        Schmidt-Kennicutt law coefficient
\item[$C_{TF}$]                                        Tully-Fisher law coefficient
\item[$E_{SN}$]                                        kinetic energy of a SNeII
\item[$\epsilon_{SN}$]                                 star formation efficiency factor 
\item[$\epsilon_{SN, fid}$]                            star formation efficiency at fiducial velocity
\item[$\epsilon_l$]                                    energy release per unit rest mass
\item[$\epsilon_m$]                                    mechanical energy release per unit rest mass 
\item[$\epsilon_n$]                                     nuclear burning efficiency per unit rest mass  
\item[${\bar \epsilon_{SN}}$]                          modified star formation factor 
\item[$\Sigma_g$]                                      gas mass surface density
\item[$\sigma_g$]                                      gas velocity dispersion
\item[$\Sigma_{cl}$]                                   cloud mass surface density
\item[$f_{cl}$]                                        gas fraction in clouds by mass
\item[$f_c$]                                           cloud volume filling factor ($= e^{-Q}$)
\item[$f_L$]                                           hot gas loading factor 
\item[$f_l$]                                           luminosity per unit mass of massive star formation
\item[$f_w$]                                           fraction accreting material that flows out in wind
\item[$H$]                                             disk scale height of gas
\item[$L_{\ast}$]                                       I-band luminosity 
\item[$\Lambda_{eff}$]                                 effective cooling function for SN bubbles
\item[$\lambda$]                                       cooling rate per atom per unit density
\item[$L_{SRF}$]                                        self-regulating feedback luminosity
\item[$L_{EDD}$]                                       Eddington lumimosity
\item[$L_{SR, fid}$]                                   self-regulating feedback luminosity at fiducial velocity dispersion
\item[$L_{mech}$]                                      mechanical luminosity
\item[$L_J$]                                           jet luminosity
\item[$\lambda_t$]                                     numerical constant  
\item[$m_{SN}$]                                        mass formed in stars per SNeII
\item[${\dot M_{out}}$]                                mass outflow rate
\item[$M_{SRF}$]                                        self-regulating feedback  mass
\item[$M_{BH}$]                                        black hole mass
\item[$M_{BE}$]                                        Bonnor-Ebert mass
\item[${\dot M_{out}}^{gal}$]                          galaxy mass outflow rate
\item[${\dot M_{\ast}}^{AGN}$]                         AGN-enhanced star formation rate
\item[${\cal M}$]                                      turbulent Mach number
\item[$\nu_h$]                                         hot phase underdensity parameter
\item[$\Omega$]                                        angular rotation of the disk
\item[$p$]                                             logarithmic slope in the black-hole mass {\it vs} galaxy plane  
\item[$p_g$]                                           gas pressure
\item[$p_{cl}$]                                        pressure in a cloud
\item[$P_{co}$]                                        cocoon pressure
\item[$Q$]                                             porosity of the interstellar medium
\item[$R$]                                             radius of disk
\item[$R_a$]                                           maximum SN bubble radius
\item[$R_0,t_0, v_0, c_0$]                             defined constants for SN bubbles
\item[$\rho_g$]                                        gas mass density
\item[$S_{cl}$]                                        covering factor of clouds
\item[$\Sigma_{\ast}$]                                 mass surface density in stars
\item[$\Sigma_{tot}$]                                  total mass surface density of the disk
\item[$\sigma_{fid}$]                                  fiducial velocity dispersion for SN driven ISM
\item[$\sigma$]                                        velocity dispersion of system
\item[$s$]                                             specific entropy
\item[$\sigma_t$]                                      PDF dispersion
\item[$t_{\ast}$]                                      gas consumption time
\item[${\bar t_{S}}$]                                  modified Salpeter time
\item[$t_{coll}$]                                      cloud collision timescale
\item[$t_a$]                                           time for SN bubble to reach maximum radius
\item[$t_c$]                                           cooling timescale of SN bubble
\item[$t_d$]                                           dynamical timescale
\item[$t_{co}$]                                        cocoon propagation timescale
\item[$\tau$]                                          Rosseland mean optical depth
\item[$t_{BH}$]                                        Black hole growth time
\item[$t_S$]                                           Salpeter time scale
\item[$\Theta_J$]                                      jet opening angle
\item[$v_c$]                                           supernova velocity at strong cooling
\item[$v_r$]                                           rotation velocity of the disk    
\item[$v_{co}$]                                        cocoon velocity
\item[$v_w$]                                           wind velocity
\item[$V_{AGN}$]                                       velocity of AGN outflow
\item[$v_J$]                                           jet velocity
\item [$\chi$]                                                interstellar cloud pressure enhancement due to self-gravity
           
\end{description}


\begin{thebibliography}{}
\bibitem[Agertz et al.(2009)]{age09} Agertz, O., Lake, G., Teyssier, R., Moore, B., Mayer, L., \& Romeo, A.~B.\ 2009, \mnras, 392, 294 
\bibitem[Alexander et al.(2008)]{ale08} Alexander, D.~M., et al.\ 2008, \aj, 135, 1968 
\bibitem[Allard et al.(2006)]{all06} Allard, E.~L., Knapen, J.~H., Peletier, R.~F., \& Sarzi, M.\ 2006, \mnras, 371, 1087 
\bibitem[Allen et al.(2006)]{alle06} Allen, S.~W., Dunn, R.~J.~H., Fabian, A.~C., Taylor, G.~B., 
\& Reynolds, C.~S.\ 2006, \mnras, 372, 21 
\bibitem[Alonso-Herrero et al.(2008)]{alo08} Alonso-Herrero, A., P{\'e}rez-Gonz{\'a}lez, P.~G., Rieke, G.~H., Alexander, D.~M., Rigby, J.~R., Papovich, C., Donley, J.~L., \& Rigopoulou, D.\ 2008, \apj, 677, 127 
\bibitem[Antonuccio-Delogu \& Silk(2008)]{ant08} Antonuccio-Delogu, V., \& Silk, J.\ 2008, \mnras, 389, 1750 
\bibitem[Babi{\'c} et al.(2007)]{bab07} Babi{\'c}, A., Miller, L., Jarvis, M.~J., Turner, T.~J., Alexander, D.~M., \& Croom, S.~M.\ 2007, \aap, 474, 755 
\bibitem[Banerjee \& Pudritz(2006)]{ban06} Banerjee, R., \& Pudritz, R.~E.\ 2006, \apj, 641, 949 
\bibitem[Barnes(2004)]{bar04} Barnes, J.~E.\ 2004, \mnras, 350, 798 
\bibitem[Begelman \& Cioffi(1989)]{beg89} Begelman, M.~C., \& Cioffi, D.~F.\ 1989, \apjl, 345, L21 
\bibitem[Blitz \& Rosolowsky(2006)]{bli06} Blitz, L., \& Rosolowsky, E.\ 2006, \apj, 650, 933 
\bibitem[Boldyrev et al.(2002)]{bol02} Boldyrev, S., Nordlund, {\AA}., \& Padoan, P.\ 2002, \apj, 573, 678 
\bibitem[Boomsma(2007)]{boo07} Boomsma, R.\ 2007, Ph.D.~Thesis, University of Groningen  
\bibitem[Bouch{\'e} et al.(2007)]{bou07} Bouch{\'e}, N., et al.\ 2007, \apj, 671, 303 
\bibitem[Bower et al.(2006)]{bow06} Bower, R.~G., Benson, A.~J., Malbon, R., Helly, J.~C., Frenk, C.~S., Baugh, C.~M., Cole, S., \& Lacey, C.~G.\ 2006, \mnras, 370, 645 
\bibitem[Cioffi et al.(1988)]{cio88} Cioffi, D.~F., McKee, C.~F., \& Bertschinger, E.\ 1988, \apj, 334, 252 
\bibitem[Croton et al.(2006)]{cro06} Croton, D.~J., et al.\ 2006, \mnras, 365, 11 
\bibitem[Di Matteo et al.(2008)]{dim08} Di Matteo, T., Colberg, J., Springel, V., Hernquist, L., 
\& Sijacki, D.\ 2008, \apj, 676, 33 
\bibitem[Fabian (1999)]{fab99} Fabian, A. \ 1999, \mnras, 308, 39
\bibitem[Feain et al.(2007)]{fea07} Feain, I.~J., Papadopoulos, P.~P., Ekers, R.~D., \& Middelberg, E.\ 2007, \apj, 662, 872 
\bibitem[Firmani \& Tutukov (1992)]{firm92} Firmani, C. \& Tutukov, A. 1992, \aa, 264, 37
\bibitem[Fragile et al.(2004)]{fra04} Fragile, P.~C., Murray, S.~D., \& Lin, D.~N.~C.\ 2004, \apj, 617, 1077 
\bibitem[Gammie et al.(1991)]{gam91} Gammie, C.~F., Ostriker, J.~P., \& Jog, C.~J.\ 1991, \apj, 378, 565 
\bibitem[Gao \& Solomon(2004)]{gao04} Gao, Y., \& Solomon, P.~M.\ 2004, \apj, 606, 271 
\bibitem[Gao et al.(2007)]{gao07} Gao, Y., Carilli, C.~L., Solomon, P.~M., \& Vanden Bout, P.~A.\ 2007, \apjl, 660, L93 
\bibitem[Hasinger et al.(2005)]{has05} Hasinger, G., Miyaji, T., \& Schmidt, M.\ 2005, \aap, 441, 417
\bibitem[Hathi et al.(2008)]{hat08} Hathi, N.~P., Malhotra, 
S., \& Rhoads, J.~E.\ 2008, \apj, 673, 686 
\bibitem[Joung \& Mac Low(2006)]{jou06} Joung, M.~K.~R., \& Mac Low, M.-M.\ 2006, \apj, 653, 1266 
\bibitem[Joung et al.(2008)]{jou08} Joung, M.~R., Mac Low, 
M.-M., \& Bryan, G.~L.\ 2008, arXiv:0811.3747 
\bibitem[Kennicutt et al.(2007)]{ken07} Kennicutt, R.~C., Jr., et al.\ 2007, \apj, 671, 333 
\bibitem[Kim \& Ostriker(2007)]{kim07} Kim, W.-T., \& Ostriker, E.~C.\ 2007, \apj, 660, 1232 
\bibitem[Komugi et al.(2005)]{kom05} Komugi, S., Sofue, Y., Nakanishi, H., Onodera, S., \& Egusa, F.\ 2005, \pasj, 57, 733 
\bibitem[Koyama \& Ostriker(2008a)]{koy08a} Koyama, H., \& Ostriker, E.~C.\ 2008, arXiv:0812.1848 
\bibitem[Koyama \& Ostriker(2008b)]{koy08b} Koyama, H., \& Ostriker, E.~C.\ 2008, arXiv:0812.1846 
\bibitem[Kriek et al.(2007)]{krie07} Kriek, M., et al.\ 2007, \apj, 669, 776 
\bibitem[Krumholz et al.(2006)]{kru06} Krumholz, M.~R., Matzner, C.~D., \& McKee, C.~F.\ 2006, \apj, 653, 361 
\bibitem[Krumholz \& McKee(2005)]{kru05} Krumholz, M.~R., \& McKee, C.~F.\ 2005, \apj, 630, 250 
\bibitem[Krumholz \& Thompson(2007)]{kru07} Krumholz, M.~R., \& Thompson, T.~A.\ 2007, \apj, 669, 289 
\bibitem[Leroy et al.(2008)]{le08} Leroy, A.~K., Walter, F., Brinks, E., Bigiel, F., de Blok, W.~J.~G., Madore, B., \& Thornley, M.~D.\ 2008, \aj, 136, 2782 
\bibitem[Lodato \& Natarajan(2006)]{lod06} Lodato, G., \& Natarajan, P.\ 2006, \mnras, 371, 1813 
\bibitem[Maiolino et al.(2007)]{mai07} Maiolino, R., et al.\ 2007, \aap, 472, L33 
\bibitem[Martin(2005)]{mar05} Martin, C.~L.\ 2005, \apj, 621, 227 
\bibitem[Martin et al.(2002)]{mar02} Martin, C.~L., Kobulnicky, H.~A., \& Heckman, T.~M.\ 2002, \apj, 574, 663 
\bibitem[Masters et al.(2008)]{mas08} Masters, K.~L., Springob, C.~M., \& Huchra, J.~P.\ 2008, \aj, 135, 1738 
\bibitem[McLure et al.(2006)]{mcl06} McLure, R.~J., Jarvis, M.~J., Targett, T.~A., Dunlop, J.~S., \& Best, P.~N.\ 2006, \mnras, 368, 1395 
\bibitem[Meurer et al.(1997)]{meu07} Meurer, G.~R., Heckman, T.~M., Lehnert, M.~D., Leitherer, C., \& Lowenthal, J.\ 1997, \aj, 114, 54 
\bibitem[Mellema et al.(2002)]{mel02} Mellema, G., Kurk, J.~D., \ Rottgering, H.~J.~A.\ 2002, \aap, 395, L13 
\bibitem[Mohanty et al.(2005)]{moh05} Mohanty, S., Jayawardhana, R., \& Basri, G.\ 2005, \apj, 626, 498 
\bibitem[Norman \& Spaans(1997)]{nor97} Norman, C.~A., \& Spaans, M.\ 1997, \apj, 480, 145 
\bibitem[Plume et al.(1997)]{plu97} Plume, R., Jaffe, D.~T., Evans, N.~J., II, Martin-Pintado, J., 
\& Gomez-Gonzalez, J.\ 1997, \apj, 476, 730 
\bibitem[Rocha-Pinto et al.(2000)]{roc00} Rocha-Pinto, H.~J., Scalo, J., Maciel, W.~J., \& Flynn, C.\ 2000, \aap, 358, 869 
\bibitem[Ryan et al.(2008)]{rya08} Ryan, R.~E., Jr., Cohen, S.~H., Windhorst, R.~A., \& Silk, J.\ 2008, \apj, 678, 751 
\bibitem[Saxton et al.(2005)]{sax05} Saxton, C.~J., Bicknell, G.~V., Sutherland, R.~S., \& Midgley, S.\ 2005, \mnras, 359, 781 
\bibitem[Schaye(2004)]{sch04} Schaye, J.\ 2004, \apj, 609, 667 
\bibitem[Shetty \& Ostriker(2008)]{she08} Shetty, R., \& Ostriker, E.~C.\ 2008, \apj, 684, 978 
\bibitem[Silk \& Rees(1998)]{sr98} Silk, J., \& Rees, M.~J.\ 1998, \aap, 331, L1 
\bibitem[Silk(2001)]{sil01} Silk, J.\ 2001, \mnras, 324, 313 
\bibitem[Silk(2005)]{sil05} Silk, J.\ 2005, \mnras, 364, 1337 
\bibitem[Silverman et al.(2008)]{silv08} Silverman, J.~D., et al.\ 2008, \apj, 679, 118 
\bibitem[Slyz et al.(2005)]{sly05} Slyz, A.~D., Devriendt, J.~E.~G., Bryan, G., \& Silk, J.\ 2005, \mnras, 356, 737 
\bibitem[Tacconi et al.(2008)]{tac08} Tacconi, L.~J., et al. 2008, \apj, 680, 246 
\bibitem[Tamburro et al.(2009)]{tam09} Tamburro, D., Rix, H.~-., Leroy, A.~K., Mac Low, M.~-., Walter, F., Kennicutt, R.~C., Brinks, E., \& de Blok, W.~J.~G.\ 2009, arXiv:0903.0183 
\bibitem[Tan(1999)]{tan99} Tan, J. 2000, \apj, 536, 173
\bibitem[Tasker \& Bryan(2006)]{tas06} Tasker, E.~J., \& Bryan, G.~L.\ 2006, \apj, 641, 878 
\bibitem[Tasker \& Tan(2008)]{tas08} Tasker, E.~J., \& Tan, J.~C.\ 2008, arXiv:0811.0207 
\bibitem[Thompson et al.(2005)]{tho05} Thompson, T.~A., Quataert, E., \& Murray, N.\ 2005, \apj, 630, 167 
\bibitem[Wada \& Norman(2002)]{wad02} Wada, K., \& Norman, C.~A.\ 2002, \apjl, 566, L21 
\bibitem[Wada \& Norman(2007)]{wad07} Wada, K., \& Norman, C.~A.\ 2007, \apj, 660, 276 
\bibitem[Walter et al. (2009)]{fab09} Walter,F., Riechers, D., Cox, P., Neri, R., Carilli, C., Bertoldi, F., Weiss, A. \& Maiolino, R., Nature, 457, 699
\bibitem[Wolfe \& Chen(2006)]{wol06} Wolfe, A.~M., \& Chen, H.-W.\ 2006, \apj, 652, 981 
\bibitem[Wolfire et al.(2003)]{wol03} Wolfire, M.~G., McKee, C.~F., Hollenbach, D., \& Tielens, A.~G.~G.~M.\ 2003, \apj, 587, 278 
\bibitem[Woo et al.(2008)]{woo08} Woo, J.-H., Treu, T., Malkan, M.~A., \& Blandford, R.~D.\ 2008,\apj, 681, 925 

\end{thebibliography}
\end{document}